\documentclass[graphicx, amsmath, floatfix, reprint, superscriptaddress, a4paper]{revtex4-1}
\usepackage{graphicx}

\begin{document}


\title{3D Simulations of Plasma Filaments in the Scrape Off Layer: \\A Comparison with Models of Reduced Dimensionality} 



\author{L. Easy}
\email[]{le590@york.ac.uk}
\affiliation{Department of Physics, University of York, Heslington, York, YO10 5DD, UK}
\affiliation{CCFE, Culham Science Centre, Abingdon, OX14 3DB, UK}

\author{F. Militello}
\affiliation{CCFE, Culham Science Centre, Abingdon, OX14 3DB, UK}

\author{J. Omotani}
\affiliation{CCFE, Culham Science Centre, Abingdon, OX14 3DB, UK}

\author{B. Dudson}
\affiliation{Department of Physics, University of York, Heslington, York, YO10 5DD, UK}

\author{\\E. Havl\'{i}\v{c}kov\'{a}}
\affiliation{CCFE, Culham Science Centre, Abingdon, OX14 3DB, UK}

\author{P. Tamain}
\affiliation{CEA, IRFM, F-13108 Saint-Paul-lez-Durance, France}

\author{V. Naulin}
\affiliation{DTU, Department of Physics, DK-2800 Kgs. Lyngby, Denmark}

\author{A. H. Nielsen}
\affiliation{DTU, Department of Physics, DK-2800 Kgs. Lyngby, Denmark}


\date{\today}

\begin{abstract}
This paper presents simulations of isolated 3D filaments in a slab geometry obtained using a newly developed 3D reduced fluid code written using the BOUT++ framework.  First, systematic scans were performed to investigate how the dynamics of a filament are affected by its amplitude, perpendicular size and parallel extent.  The perpendicular size of the filament was found to have a strong influence on its motions, as it determined the relative importance of parallel currents to polarisation and viscous currents, whilst drift-wave instabilities were observed if the initial amplitude of the blob was increased sufficiently.  

Next, the 3D simulations were compared to 2D simulations using different parallel closures; namely, the \textit{sheath dissipation} closure, which neglects parallel gradients, and the \textit{vorticity advection} closure, which neglects the influence of parallel currents.  The vorticity advection closure was found to not replicate the 3D perpendicular dynamics and overestimated the initial radial acceleration of all the filaments studied.  In contrast, a more satisfactory comparison with the sheath dissipation closure was obtained, even in the presence of significant parallel gradients, where the closure is no longer valid.  Specifically it captured the contrasting dynamics of filaments with different perpendicular sizes that were observed in the 3D simulations which the vorticity advection closure failed to replicate.  However, neither closure successfully replicated the \textit{Boltzmann spinning} effects and associated poloidal drift of the blob that was observed in the 3D simulations. 

Although the sheath dissipation closure was concluded to be more successful in replicating the 3D dynamics, it is emphasised that the vorticity closure may still be relevant for situations where the parallel current is inhibited from closing through the sheath due to effects such as strong magnetic shear around X points or increased resistivity near the targets.  
\end{abstract}


\maketitle 



%
%

%

\section{Introduction}
\label{sec:into}
A key characteristic of turbulence observed in the Scrape Off Layer (SOL) of magnetic confinement devices is the presence of coherent field aligned plasma structures, called \textit{filaments} or \textit{blobs}, that are significantly more dense and hot than their surrounding plasma and strongly localised in the drift-plane perpendicular to the equilibrium magnetic field \cite{Zweben:2007dp}.  In the SOL of both L-mode and inter-ELM H-mode plasmas, particle transport appears to be dominated by non-local, rather than diffusive processes \cite{Garcia:2007hh}, with measurements from the DIII-D tokamak indicating that approximately 50\% of cross field particle transport can be attributed to the advection of filaments in both regimes \cite{Boedo:2003fe}.  The motions of filaments therefore have a significant influence on the particle (and possibly to a lesser extent heat) fluxes to the divertor and first wall, and as such, a full understanding of their dynamics is essential for the successful operation of future fusion experiments and reactors.  

The basic mechanism of radial filament advection, first proposed by Krashenninikov \cite{Krashenninikov:2001uc} can be understood by considering the balance of perpendicular and parallel currents that flow through an isolated SOL filament.  Here, as for the remainder of this paper, the terms \textit{perpendicular} and \textit{parallel} are used with respect to the magnetic field direction.  As a result of the filament's strong cross field pressure gradients, diamagnetic currents will flow in the perpendicular plane.  In a uniform, straight magnetic field, these currents will form closed circuits along the filament's density contours and no dynamics of note occur (in the absence of a neutral species).  However, in the region of the outboard midplane of a tokamak, where the magnetic field is predominantly toroidal, magnetic gradients and curvature are present, which for a given pressure gradient act to enhance the diamagnetic currents at larger radial distances.  In this case, the diamagnetic current field is no longer divergence free, which necessitates additional currents in the filament to satisfy current continuity.  The circuit can be completed directly in the drift-plane via perpendicular polarisation currents, or through parallel currents that are closed elsewhere along the field line or through the sheath.  The polarisation current path usually leads to the formation of a broadly dipolar electrostatic potential field in the perpendicular plane of the filament, and through 
$E \times B$ motions, this potential structure corresponds to a pair of counter rotating vortices that act to advect it radially outwards.  The detailed motions of the filament are dictated by the exact structure of the potential field, which in turn is determined by the strength of the diamagnetic current drive and the effective resistances of the parallel and polarisation current paths.  

Until recently, SOL turbulence and filament theory and simulations have been predominantly two-dimensional in nature, using simple models for the parallel direction whilst retaining the full perpendicular dynamics.  Reference \cite{Krasheninnikov:2008fw} provides a comprehensive review of these works, and of the various parallel closures that have been used.  One such closure that is commonly used in the literature for isolated filament studies \cite{Bian:2003bq, Yu:2003eoa, Garcia:2006dt, Yu:2006ci} and SOL turbulence simulations \cite{Russell:2009dx} is the \textit{Sheath Dissipation} model.  This closure assumes the filament to be \textit{sheath connected}, extending from target to target with negligible parallel gradients of density and electrostatic potential.  This allows the dynamics of the filament to be integrated and averaged over the parallel direction, with the parallel current effects represented by sheath current boundary conditions.  

Filament dynamics using this model have been shown to be largely dependent on the perpendicular density gradients present in the structure, and scaling laws have been obtained relating a filament's radial velocity to its perpendicular length scale, $\delta_\perp$ \cite{Yu:2003eoa, Theiler:2009fh}.  For filaments much below a critical size (see Equation \eqref{eq:d*}), the diamagnetic currents are predominately closed by the polarisation currents and the radial velocity is predicted to scale like $v_{b} \propto \sqrt{\delta_\perp}$.  In contrast, current continuity is primarily achieved through the sheath currents for blobs significantly larger than this critical length, and the radial velocity is estimated to decrease like $v_{b} \propto 1/\delta_\perp^2$.  In each of these two cases, the behaviours of the filament motions are qualitatively different.  Filaments in the small blob regime rapidly form a mushroom-like structure through interchange motions and subsequently lose coherence due to a combination of secondary Kelvin Helmholtz instabilities, collisional diffusion and stretching of the leading front.  In the large blob regime on the other hand, a finger-like structure is extruded from the initial filament that advects radially outwards and whose front also undergoes interchange motions.  Most significantly, filaments that sit between these two regimes exhibit the most coherent motion and retain their blob-like structure long into their evolution.  These contrasting behaviours are illustrated in Figure 4 of Reference \cite{Angus:2012fp}.  However, experimental measurements have shown that filaments tend to be localised or \textit{ballooned} around the outboard midplane \cite{Terry:2003gz} and therefore the neglect of parallel density gradients in the sheath dissipation model may not be well justified.  Despite this, turbulence simulations using sheath dissipation have found favourable comparisons with experimental results from NSTX \cite{Russell:2011gx, Myra:2011io}.  

One alternative closure \cite{Fundamenski:2007gk} seeks to better represent the ballooned nature of filaments by neglecting parallel currents rather than parallel gradients.  In this \textit{vorticity advection} closure, the radial velocity of filaments can be estimated to scale as $v_{b} \propto \sqrt{\delta_\perp}$ for all $\delta_\perp$ of interest and in contrast to the sheath dissipation model, distinct propagation regimes dependent on the perpendicular size of the filament do not exist.  The closure has been extensively used for SOL turbulence studies \cite{Garcia:2004km, Garcia:2005eo}, and like the sheath dissipation model, has obtained good agreement with experimental measurements of SOL profiles and turbulence statistics from a variety of experimental devices \cite{Fundamenski:2007gk, Garcia:2005ki, Militello:2013cl}.  With both of these 2D representations finding agreement with experiments, it unclear which model is best suited for SOL studies and systematic comparisons with 3D simulations are therefore required.  

It is only in recent years that 3D numerical investigations of isolated filaments have been performed and in References \cite{ Angus:2012fp, Angus:2012gk, Angus:2012cv}, comparisons were made against 2D simulations employing the sheath dissipation model.  These works demonstrated that the inclusion of 3D physics can lead to drift-wave turbulence that dissipates density on much faster time-scales compared to 2D models, and which therefore produces a large reduction in radial particle transport.  Moreover, in the presence of density gradients along field lines, the parallel potential field was observed to obey a Boltzmann-like response, which caused density and potential to become in phase in the perpendicular drift plane.  Through $\boldsymbol{E}\times\boldsymbol{B}$ motions, this alignment corresponded to the filament rotating in the drift-plane, and this \textit{Boltzmann spinning} again inhibited the radial motions compared to 2D simulations \cite{Angus:2012cv}.  
Numerical studies have also been performed using the magnetic geometry of a Simple Magnetised Torus (SMT) \cite{Halpern:2014ip}, and whilst no direct comparisons with 2D closures were made, the perpendicular length scale dependence predicted by the sheath dissipation closure was observed.  In addition, simulations investigating the effects of realistic magnetic geometry have shown that the dynamics of a filament at the midplane can become independent of the divertor region and hence the parallel boundary conditions at the sheath \cite{Walkden:2013dm}.  

With the exception of ~\cite{Halpern:2014ip}, in each of these 3D works, the parallel ion velocity was assumed to be negligible throughout the domain with the justification that the perpendicular motions and parallel electron dynamics occur on significantly faster time scales than that of parallel density transport to the sheath \cite{Angus:2012fp}.  Whilst such an approach may be valid to qualitatively highlight features of 3D filament dynamics, the balance between parallel and perpendicular transport is critical in the SOL, and clearly parallel density variations will quickly develop in blobs seeded without such gradients under the influence of sonic sheath boundary conditions.  Moreover, given the vorticity advection model's emphasis on parallel advection, comparisons of this closure with 3D simulations without these effects would not be consistent.  

In this paper, we present the first results from a newly developed 3D non-linear code for SOL simulations implemented using the BOUT++ framework \cite{Dudson:2009ig}.  The electrostatic fluid model employed retains parallel ion dynamics as well as the effects of finite electron inertia, which were also absent in the many of the 3D simulations previously discussed \cite{ Angus:2012fp, Angus:2012gk, Angus:2012cv, Walkden:2013dm}.  This code has been used to investigate the dynamics of isolated filaments and in particular their dependence on the filament's initial amplitude, perpendicular length scale and parallel extent, in addition to the strength of magnetic curvature.  The parallel dynamics of the code have been validated against both analytical shock propagation theory results and the 1D SOL code SOLF1D, \cite{Havlickova:2011dc}, whilst direct comparisons have been made against 2D simulations utilising the sheath dissipation and vorticity advection model to determine which 2D model best captures the dynamics observed in 3D simulations.  

The remainder of this paper is organised as follows.  In Section \ref{sec:eqns}, the governing equations of the 3D model are discussed and from which, brief derivations of both 2D closures are provided.  Next, Section \ref{sec:Numerics} outlines the numerical implementation of the simulations before Section \ref{sec:results} presents and discusses the results of the Simulations.  Specifically, Section \ref{sec:bg_fields} describes and validates the source driven backgrounds used for the 3D filament simulations, the dynamics of which are investigated in Section \ref{sec:3Dsims} through a series of parameter scans.  Following this, Section \ref{sec:1D3D} describes the validation of the parallel dynamics against SOLF1D and analytic theory, before Section \ref{sec:2D3D} compares the 3D results against those produced by 2D simulations.  Finally the conclusions of the paper are summarised in Section \ref{sec:conclusions}.  
\section{Governing Equations}
\label{sec:eqns}
\subsection{3D Model}
\label{sec:3Dmodel}
The 3D results presented in this work have been obtained using an electrostatic drift-fluid model which assumes singly charged cold ions and isothermal electrons.  A slab geometry has been used with a uniform magnetic field $\textbf{B}=B\boldsymbol{\hat{z}}$, whilst the effects of magnetic curvature and gradients have been represented through additional terms in the evolution equations.  The $x$ and $y$ coordinates in this geometry represent the effective radial and poloidal directions respectively.  Throughout this paper, a Bohm normalisation has been employed, with time and length scales normalised to the ion gyro-frequency, $\Omega_i = eB/m_i$, and the hybrid gyro-radius $\rho_s = c_s/\Omega_i$ respectively, whilst the electrostatic potential, $\phi$ has been normalised to $T_e/e$. Here $e$ is the elementary unit charge, $m_i$ is the ion mass, $c_s = \sqrt{T_e/m_e}$ is the sound speed, $T_e$ is the electron temperature in Joules and $m_e$ is the mass of an electron.  In addition, the plasma density has been normalised to a characteristic background SOL density, $n_{SOL}$.  The resulting non dimensional evolution equations for plasma density, $n$, vorticity, $\Omega = \nabla_\perp^2\phi$, parallel ion velocity, $U$, and parallel electron velocity, $V$, are:
\begin{align} 
\dfrac{dn}{dt}
& = - \nabla_\parallel\left( n V\right) 
+ ng\dfrac{\partial \phi}{\partial y} 
- g \dfrac{\partial n}{\partial y} 
+ D_n\nabla_\perp^2 n 
+ S_n,
\label{eq:n}
\\
\dfrac{d\Omega}{dt} 
&= -U\nabla_\parallel\Omega 
+ \dfrac{1}{n}\nabla_\parallel J_\parallel 
- \dfrac{g}{n} \dfrac{\partial n}{\partial y}
+ \mu_i\nabla_\perp^2\Omega, 
\label{eq:vort}
\\
\dfrac{dU}{d t} 
&= -U\nabla_\parallel U
- \nabla_\parallel\phi
- \dfrac{\nu_{\parallel}}{\mu n}J_\parallel
- \dfrac{S_nU}{n},
\label{eq:U}
\\
\dfrac{dV}{dt} 
&= - V\nabla_\parallel V
+ \mu\nabla_\parallel\phi
- \dfrac{\mu}{n}\nabla_\parallel n
+ \dfrac{\nu_{\parallel}}{n}J_\parallel
- \dfrac{S_nV}{n}.
\label{eq:V} 
\end{align}
Here, $\frac{d}{dt} = (\frac{\partial}{\partial t} + \boldsymbol{\hat{z}}\times\nabla\phi\cdot\nabla)$, $J_\parallel = n(U-V)$ is the normalised parallel current density, $S_n$ is a particle source, $\mu = m_i/m_e$ is the ratio of ion to electron masses, and $\nu_\parallel = \nu_{ei}/2\Omega_i$ is the normalised electron - ion collision frequency where $\nu_{ei} = ne^4\ln \Lambda/3m_e^{1/2}\epsilon_0^2 (2\pi T_e)^{3/2}$, $D_n = (1+1.3q^2)(1+\frac{T_i}{T_e})\nu_{ei}/(\mu \Omega_i)$ is the normalised particle perpendicular diffusivity, $\mu_i = \frac{3}{4}(1+1.6q^2)\sqrt{m_e/T_i}\rho_s\nu_{ei}$ is the normalised ion perpendicular viscosity, $q$ is the tokamak safety factor, $T_i$ is the ion temperature in Joules and $\epsilon_0$ is the permittivity of free space.  $D_n$ and $\mu_i$ are defined as in \cite{Fundamenski:2007gk}, and it is noted that whilst cold ions are assumed in this model, finite ion temperatures are retained for the calculation of these dissipative parameters.  

Equation \eqref{eq:n} is a statement of density conservation and whilst the first three terms on its RHS are often neglected for filament studies \cite{Bian:2003bq,  Garcia:2006dt, Angus:2012gk, Angus:2012fp} as they are of lower order than the LHS, they are retained here for completeness.  Consideration of current continuity and application of the Boussinesq approximation produce Equation \eqref{eq:vort} and its LHS and the first term on its RHS are of the divergence of the ion polarisation current density, $ \boldsymbol{J_{P}}$, divided by $n$.  The second and last terms on its RHS are respectively the divergences of the parallel current density, $J_\parallel$, and the current density arising from viscous effects, $\boldsymbol{J_{\mu_i}}$, also divided by $n$.  The remaining term on the RHS of Equation~\eqref{eq:vort} corresponds to the divergence of the electron diamagnetic current density, $ \boldsymbol{J_{D}} = \boldsymbol{\hat{b}}\times\nabla n/B$, divided by $n$, through which the $\nabla B$ and curvature effects force the other currents in the system.  The strength of such a drive has been represented through the parameter $g$, which in a tokamak can be approximated to be $g = 2\rho_s/R_c$, where $R_c$ is the dimensional radius of curvature.  Equations \eqref{eq:U} and \eqref{eq:V} are the parallel momentum equations for each particle species and the last term on the RHS in each ensures that momentum in the system is conserved in the presence of a particle source.  At the location of the targets, $z = \pm L_\parallel$, where $L_\parallel = \ell_\parallel/\rho_s$ is the normalised midplane to target distance, the velocity fields evolved by these two equations must satisfy standard sheath boundary conditions \cite{Stangeby:2008ta}:
\begin{align}
\left. U \right|_{z=\pm L_\parallel} &= \pm 1 \label{eq:Ubnd}, \\ 
\left. V \right|_{z=\pm L_\parallel} &= \pm \exp\left(- \phi\right),\label{eq:Vbnd}
\end{align}
where $\phi$ is defined with respect to the plasma's floating potential.  Equation \eqref{eq:Ubnd} is the Bohm sheath criterion for ions, specifying a sonic velocity at the entrance to the sheath, whilst Equation \eqref{eq:Vbnd} specifies that electrons travel slower (faster) than the ion sound speed into the sheath for $\phi < 0$ ($\phi > 0$), thereby transiently allowing currents to flow into or out of sheath.  The remaining boundary conditions are described in Section \ref{sec:Numerics}.  

\subsection{2D Closures}
\label{sec:2Dclosures}
In order to form a closed system of 2D equations from this 3D model, assumptions must be made regarding the parallel dynamics.  As described previously, the sheath dissipation closure assumes the filament to be sheath connected, with negligible gradients of density and potential in the parallel direction.  By application of a linearised form of the sheath boundary conditions above, Equations and \eqref{eq:n} and \eqref{eq:vort} can be integrated and averaged along the parallel direction to produce:
\begin{align} 
& \dfrac{dn}{d t} 
= \dfrac{n \phi}{L_\parallel} 
-\dfrac{\left( n - n_0\right) }{L_\parallel} 
+ ng\dfrac{\partial \phi}{\partial y}
- g \dfrac{\partial n}{\partial y}
+ D_n\nabla_\perp^2 n, 
\label{eq:n_SD}
\\
& \dfrac{d\Omega}{dt} 
= \dfrac{\phi}{L_\parallel}
- \dfrac{g}{n} \dfrac{\partial n}{\partial y}
+ \mu_i\nabla_\perp^2 \Omega
 \label{eq:vort_SD}.
\end{align}
Here, $n_0$ is the constant SOL background density, and it has been additionally assumed in obtaining Equations \eqref{eq:n_SD} and \eqref{eq:vort_SD} that the parallel integral value of $S_n$ is equal to $2n_0$.  These sheath dissipation model equations describe the parallel averaged motions of density and vorticity in the system.   

Viscous effects are small in SOL plasmas, and so it is generally assumed in the sheath dissipation model that the non divergence free diamagnetic currents are essentially closed through a combination of polarisation and sheath currents.  It has been demonstrated that the perpendicular length scale of the filament, $\delta_\perp$, controls which of these two currents paths is dominant in satisfying current continuity \cite{Yu:2003eoa}.  Following the scaling arguments outlined in \cite{Theiler:2009fh} and \cite{Angus:2012fp}, it can be estimated that for a filament with peak density perturbation $\delta n$, the polarisation and sheath current terms in Equation \eqref{eq:vort_SD} are of the same order when $\delta_\perp = \delta_*$, where
\begin{equation} \label{eq:d*}
\delta_* = \left( \dfrac{gL_\parallel^2}{2}\dfrac{\delta n}{n_0+\delta n} \right) ^{1/5},
\end{equation}
and that the typical radial velocity of the blob, $v_{b}$ when it is travelling coherently will be approximately:
\begin{equation} \label{eq:vblob}
v_{b} = {v_b}_0  
 \dfrac{\left( \dfrac{\delta_\perp}{\delta_*}\right) ^{1/2}}
{1 + \sqrt{\dfrac{1}{2}\dfrac{\delta n}{n_0+\delta n}}\left(\dfrac{\delta_\perp}{\delta_*} \right)^{5/2} },
\end{equation}
where ${v_b}_0 = (\frac{1}{2}g^6L_\parallel^2(\frac{\delta n}{n_0+\delta n} )^{11} )^{1/10}$.  

Alternatively, the vorticity advection model considers a drift plane in the region of the outboard midplane, neglects parallel currents, and estimates parallel advection terms to be $U\nabla_\parallel = V\nabla_\parallel \approx 0.5L_{b}$, where $L_b$ is a characteristic parallel length scale of the filament to produce:
\begin{align} 
& \dfrac{dn}{dt} 
= -\dfrac{ \left( n - n_0\right) }{2 L_{b}} 
+ ng\dfrac{\partial \phi}{\partial y} 
- g \dfrac{\partial n}{\partial y}
+ D_n\nabla_\perp^2 n, 
 \label{eq:n_VA}
 \\
&\dfrac{d\Omega}{dt} 
= - \dfrac{\Omega}{2L_{b}}
- \dfrac{g}{n} \dfrac{\partial n}{\partial y}
+ \mu_i\nabla_\perp^2\Omega.  
\label{eq:vort_VA}
\end{align}
It is pertinent to note upon inspection of the vorticity equations of these 2D models that the sheath dissipation term in Equation \eqref{eq:vort_SD} preferentially damps larger potential scale lengths, whilst the vorticity advection term in Equation \eqref{eq:vort_VA} acts on all potential scale lengths equally.  

\section{Numerical Implementation}
\label{sec:Numerics}
With the exception of the benchmark comparison cases in Section \ref{sec:1D3D}, all of the simulations presented in this paper were completed using the BOUT++ framework.  For numerical stability, staggered grids were employed in the parallel direction, alongside a first order upwinding scheme for the parallel advection derivatives and an Arakawa scheme \cite{Arakawa:1966gs} for the perpendicular $E\times B$ advective derivatives.  All other derivatives were calculated using second order central differencing.  

In the 3D simulations, only half the parallel domain has been simulated for computational efficiency, with symmetry boundary conditions employed at the $z = 0$ parallel boundary.  At the other parallel boundary, $z = L_\parallel$, sheath boundary conditions given by \eqref{eq:Ubnd} and \eqref{eq:Vbnd} were used, whilst no boundary conditions were set on the remaining variables to prevent over-constraining the system.  In the perpendicular plane, the $y$ direction is periodic, whilst at the $x$ boundaries, $\Omega$ was fixed to zero, and zero gradients were imposed on $n$, $U$ and $V$.  

The remaining perpendicular boundary condition for $\phi$ was set as to obtain the 1D (variation only in the parallel direction) steady state fields described in Section \ref{sec:bg_fields}.  This was achieved by fixing the $x$ boundaries of $\phi$  to the parallel profile of its resulting equilibrium field.  However, as this boundary condition could not be determined \textit{a priori}, it was obtained by iteratively running the simulation until achieved a steady state equilibrium and updating the $\phi$ boundary condition to its parallel profile along the centre of the domain until a time invariant 1D system was produced.  

Identical boundary conditions were used for the 2D simulations as in the perpendicular plane of the 3D simulations, with the exception of the $x$ boundaries of the $\phi$ field, which were fixed to zero.  Systematic scans were performed to ensure that all of the simulations were sufficiently resolved and that the filaments' dynamics were not influenced by their proximity to the perpendicular boundary conditions.  The resulting perpendicular domain size and resolution were scaled according to the perpendicular size and peak amplitude of the initial seeded filament and are specified alongside the results presented in Section \ref{sec:results}.  

\section{Results}
\label{sec:results}
MAST relevant parameters have been used for these numerical investigations, with $T_e = T_i = 40$eV, $B_0=0.5$T, $n_{0}=0.8\times10^{13}$cm$^{-3}$, $\ell_\parallel=10$m, $R_c = 1.5$m, $q = 7$ and Deuterium ions \cite{Militello:2011gm}.  These primary parameters correspond to $\rho_s = 1.8$mm and produce the following non-dimensional parameters that have been used as the basis for the filament simulations presented in this work: $g = 0.0025$, $\mu = 3645$, $\nu_\parallel = 0.02$, $D_n = 0.0015$,  $\mu_i = 0.04$ and $L_\parallel = 5500 $.  

\subsection{Background Equilibrium}
\label{sec:bg_fields}
\begin{figure}
\centering
\includegraphics[trim = 0mm 0mm 0mm 0mm, clip, width = 8.5cm]{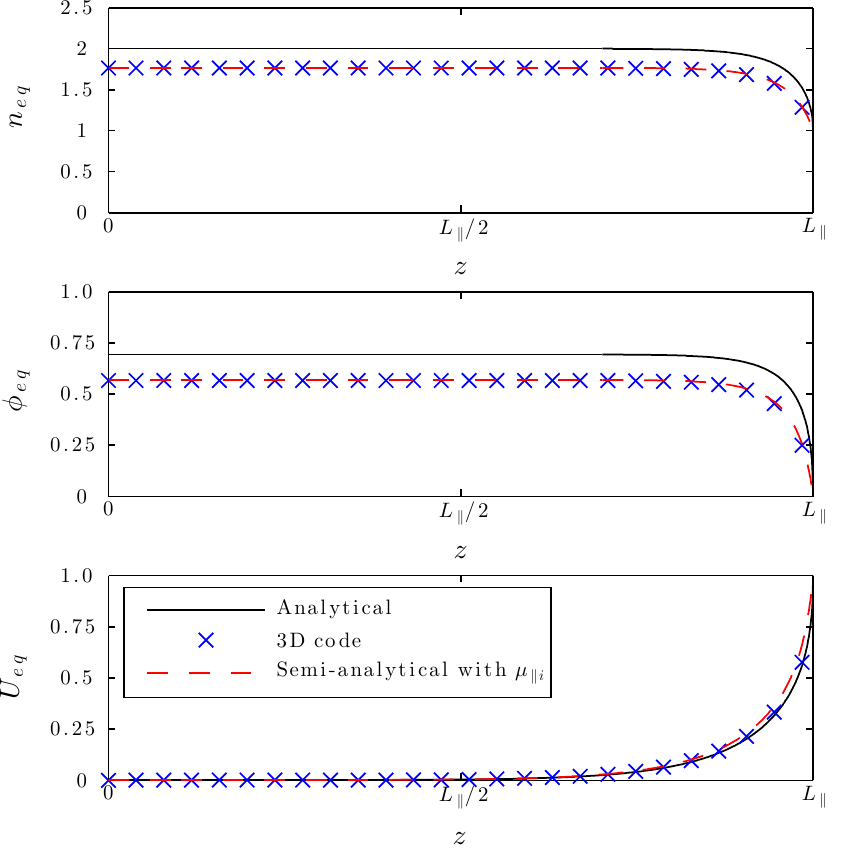}
\caption{Validation of the steady state equilibrium obtained using the 3D code against 1D analytical fields without parallel dissipative effects and semi-analytical results with parallel ion viscosity included.  }
\label{fig:eqs}	 
\end{figure}

In order to isolate the dynamics of single filaments from the evolution of the equilibrium, steady state background fields with variation only in the parallel direction were required, onto which density perturbations could be seeded.  This was achieved by using the following density source:
\begin{equation}
S_n = \dfrac{10\exp\left(10z/L_\parallel \right) }{L_\parallel\left( \exp\left(10 \right) - 1 \right) }.
\end{equation}
In the absence of perpendicular variation, the equilibrium fields, denoted by the suffix $eq$, arising from Equations \eqref{eq:n} to \eqref{eq:V} can be obtained analytically.  $\Omega_{eq}$ is necessarily zero, whilst the steady state density and velocity fields are given by:
\begin{equation}
n_{eq} = \dfrac{1}{U_{eq}}\int^z_0 S_n dz',  \label{eq:n_eq}\\
\end{equation}
\begin{equation}
U_{eq} =  V_{eq} = \dfrac{2+1/\mu - \sqrt{\left( 2+1/\mu\right) ^2 - 4\left( 1+1/\mu\right) \alpha^2}}{2\left( 1+1/\mu\right)\alpha}, \label{eq:U_eq}
\end{equation}
where $\alpha = \int_0^z S_n dz'/\int_0^{L_\parallel} S_n dz$.  The remaining equilibrium field, $\phi_{eq}$, can be obtained by assuming the Boltzmann relation $\phi = \ln n$.  These results were used to validate the steady state fields obtained using the 3D filament code, and a comparison between the two is shown in Figure \ref{fig:eqs}. %
Whilst good agreement is found for $U_{eq}$, the 3D code's upstream values of $n_{eq}$ and $\phi_{eq}$ are systematically smaller than their corresponding analytical values.  We attribute this discrepancy to be due to the presence of numerical dissipation in the 3D code, primarily resulting from the upwinding scheme used for the parallel advection derivatives.  This mechanism is demonstrated by the third data series in Figure \ref{fig:eqs}, which plots semi-analytical results for the equilibrium fields obtained with a parallel viscosity term $+ \mu_{i\parallel} \nabla_\parallel^2 U$ included on the RHS of Equation \eqref{eq:U}.  These fields are an extension of Equations \eqref{eq:n_eq} and \eqref{eq:U_eq} and were acquired using numerical integration methods.  It is clear that inclusion of such dissipative effects act to reduce the upstream $n_{eq}$ and $\phi_{eq}$ values, whilst having little effect on the $U_{eq}$ and therefore agreement with the 3D code's results is found.  In the remainder of this work, the density fields have been rescaled such that $n_{eq}$ at the midplane ($z = 0$) is unity, corresponding to the dimensional characteristic SOL density $n_{SOL}$.  

\subsection{3D Filament Dynamics}
\label{sec:3Dsims}

The 3D filaments simulations presented in this study were initialised by seeding a density perturbation $n_{b} = n - n_{eq}$ onto the background.  In line with previous works, a Gaussian profile was used in the perpendicular plane, whilst in the parallel direction a step function has been employed: 
\begin{equation}
 n_{b}(t =0) = \left\{ 
  	\begin{array}{l l}
    	\delta n\exp\left(-\dfrac{x^2+y^2}{\delta_\perp^2}\right) & \quad z \leq L_b\\
    	0 & \quad z > L_b
  	\end{array} \right., 
\end{equation}
where $\delta n$ is the filament's peak amplitude, whilst $\delta_\perp$ and $L_{b}$ respectively define its perpendicular size and parallel extent.  

\begin{figure}
\centering
\includegraphics[trim = 0mm 0mm 0mm 0mm, clip, width = 8.5cm]{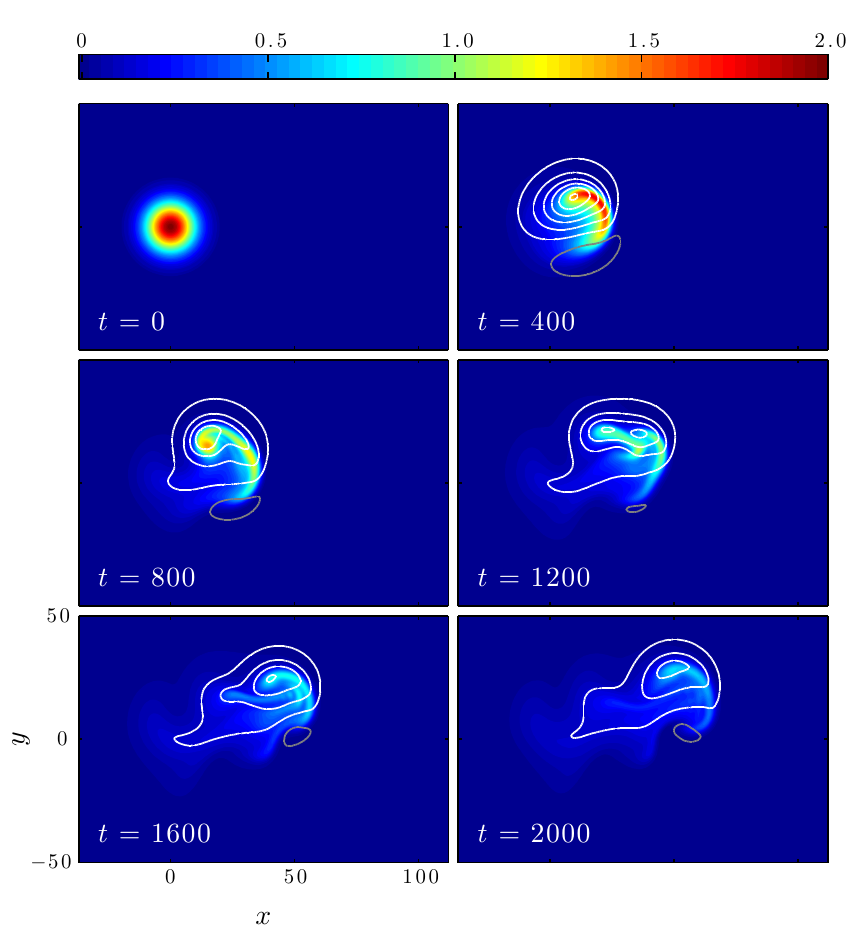}
\caption{Evolution of the density perturbation, $n_b$ (colour map) and potential, $\phi$ (contour lines), at the midplane ($z = 0$) of a filament initialised with $\delta n = 2n_0$, $L_b = L_\parallel/2$ and $\delta_\perp = 10$.  Grey and white contour lines respectively indicate values of $\phi$ less than and greater than the equilibrium value on the drift plane. }
\label{fig:delta10_conts}  	 
\end{figure}

\begin{figure}
\centering
\includegraphics[trim = 0mm 0mm 0mm 0mm, clip, width = 8.5cm]{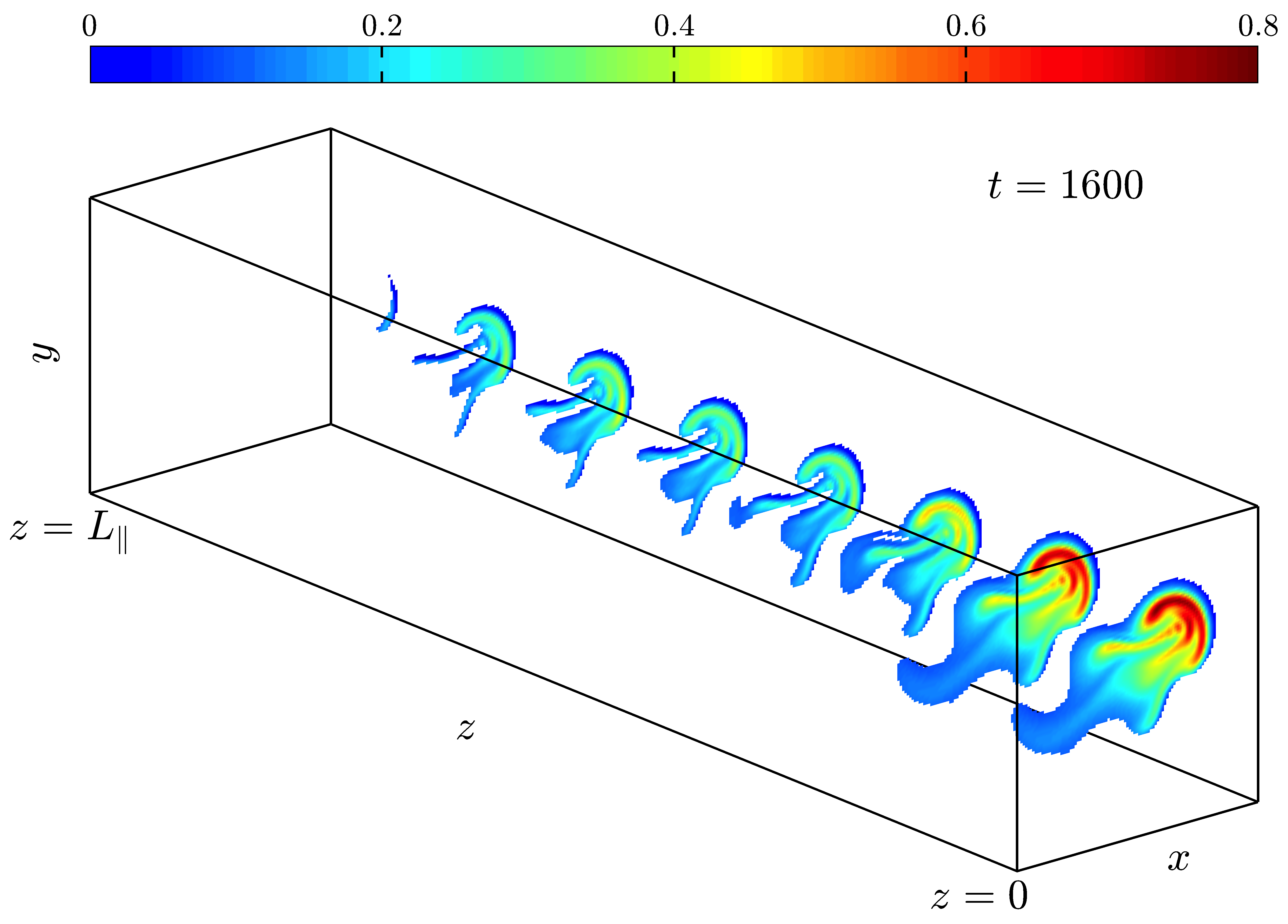}
\caption{3D structure of the density perturbation, $n_b$, of a filament seeded with $\delta n = 2n_0$, $L_b = L_\parallel/2$ and $\delta_\perp = 10$ at $t = $ 1600. }
\label{fig:3Dexample}	 
\end{figure}

For the purpose of investigating the influence of each of these parameters, the dynamics of a reference case consisting of $\delta n = 2n_0$, $L_{b} = L_\parallel/2$ and $\delta_\perp = 10$, corresponding to $\delta_\perp \approx 1.3 \delta_*$, according to Equation \eqref{eq:d*}, is presented first.  For this case, and for remainder of this paper unless stated otherwise, a perpendicular domain of $L_x \times L_y = 15\delta_\perp \times 10\delta_\perp$ and a grid resolution of $N_x \times N_y \times N_z  = 192 \times 128 \times 16$ has been employed, whilst the filament has been initialised off centre in the $x$ direction to allow for its later motions.  These dynamics are shown in Figure \ref{fig:delta10_conts}, %
which plots the time evolution of $n_b$ (colour map) and $\phi$ (contour lines) at the midplane ($z = 0$).  In this figure, as in all density and potential contour plots in this paper, white and grey contour lines respectively correspond to values of $\phi$ above and below $\phi_{eq}$ on the drift plane shown.  The midplane dynamics plotted in this figure are broadly representative of the entire filament, as demonstrated by Figure \ref{fig:3Dexample}, %
which shows its 3D density structure at a point in time late in the filament's motion. 

The interchange behaviours observed are qualitatively similar to those found in the previous 2D filament simulations discussed in Section \ref{sec:into}.  The filament quickly develops a potential structure which acts to advect the density radially outwards, leading to a steep blob front early in the simulation, and a mushroom-like structure at later times.  The most notable departure from previous 2D simulations is that $\phi$ is subject to a Boltzmann response which acts to align it with $n$ due to the $\nabla_\parallel(\phi - \ln n)$ drive in Equation \eqref{eq:V}.  Such a response has been found in previous 3D simulations \cite{Angus:2012fp, Walkden:2013dm}, and can be observed in Figure \ref{fig:delta10_conts} to make the magnitude of the top positive $\phi$ lobe of the dipole stronger than that of the negative lobe.  The $\phi$ field can thus be seen as a superposition of a weaker dipolar and a stronger monopolar components.  Through $\boldsymbol{E}\times \boldsymbol{B}$ motions, the alignment of $\phi$ and $n$ causes the filament to spin anticlockwise in the perpendicular plane (when the magnetic field acts away from the observer), and this \textit{Boltzmann spinning} acts to rotate the dipolar $\phi$ component and produce a net upward displacement in the poloidal or $y$ direction, which is clearly visible.  The asymmetry between the top and bottom $n$ lobes is produced as a result of the monopolar $\phi$ component, which produces $\boldsymbol{E}\times \boldsymbol{B}$ velocities that act with the dipolar component's velocities in the top lobe and against in the bottom lobe.  In fact, the top lobe is sheared off from the leading front just after $t = 800$ due to its spinning motions, before being rapidly advected through the filament's wake to rejoin the main front at around $t = 1200$.  It is noted that the unstable drift-waves dynamics which were the primary subject of references \cite{Angus:2012gk} and \cite{Angus:2012fp} were not present in this simulation due to dissipative and viscous effects.  As the dissipative parameters used are physically justified \cite{Fundamenski:2007gk}, this indicates that filaments may exist in the MAST SOL that are drift-wave stable.  

\begin{figure}
\centering
\includegraphics[trim = 0mm 0mm 0mm 0mm, clip, width = 8.5cm]{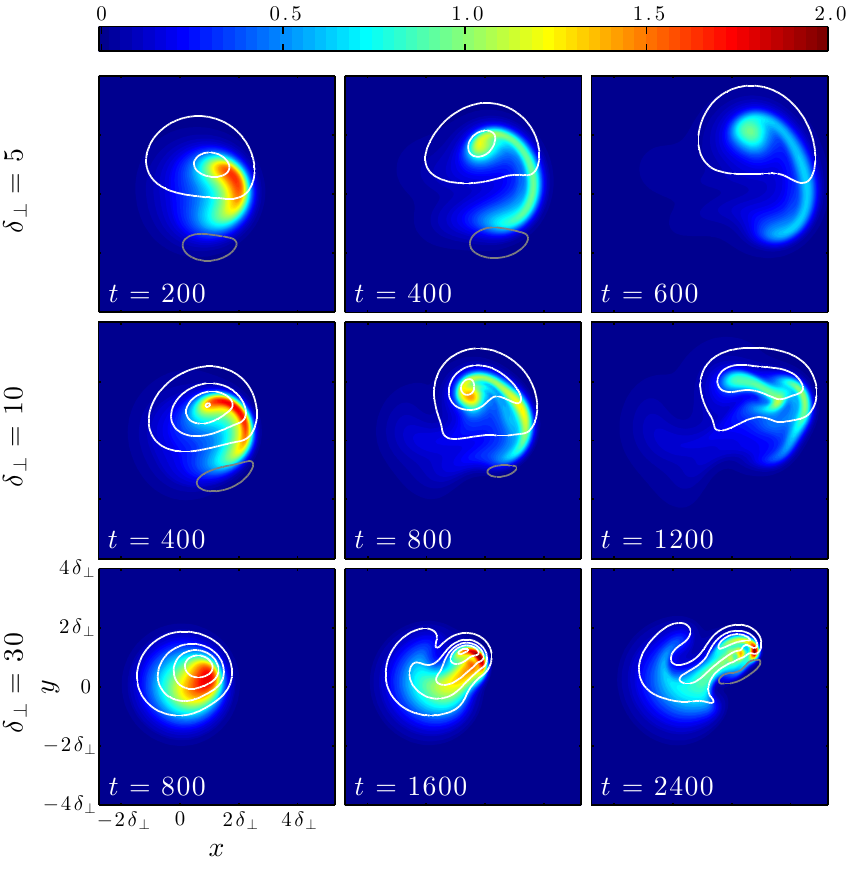}
\caption{Contrasting midplane evolution of $n_b$ (colour map) and $\phi$ (contour lines) of $\delta n = 2n_0$, $L_b = L_\parallel/2$ filaments initialised with $\delta_\perp = 5$, $\delta_\perp = 10$ and $\delta_\perp = 30$.  Grey and white contour lines respectively indicate values of $\phi$ less than and greater than the equilibrium value on the drift plane.}
\label{fig:deltascan_conts}	 
\end{figure}

The dependence of the filament's subsequent motions on its initial perpendicular size is shown in Figure \ref{fig:deltascan_conts}, %
which shows the contrasting time evolution of the $n_b$ and $\phi$ fields at the midplane of $L_b = L_\parallel/2$, $\delta n = 2n_0$ filaments with varying $\delta_\perp$.  The smallest filament, $\delta_\perp = 5$, develops a mushroom cap structure approximately twice as quickly as the reference $\delta_\perp = 10$ case, before losing coherence due to a combination of collisional diffusion and the stretching of its leading front by the counter rotating vortices.  On the other hand, the largest filament, $\delta_\perp = 30$, displays qualitatively different dynamics to the two smaller filaments.  Instead of initially moving largely as a single coherent body, it expels a finger like structure, whose front undergoes similar dynamics to those of the reference filament.  It is pertinent to note here that these morphological differences in the filaments' evolution for different $\delta_\perp$ are similar to those that arise in 2D sheath dissipation simulations, despite the presence of parallel gradients.

\begin{figure}
\centering
\includegraphics[trim = 0mm 0mm 0mm 0mm, clip, width = 8.5cm]{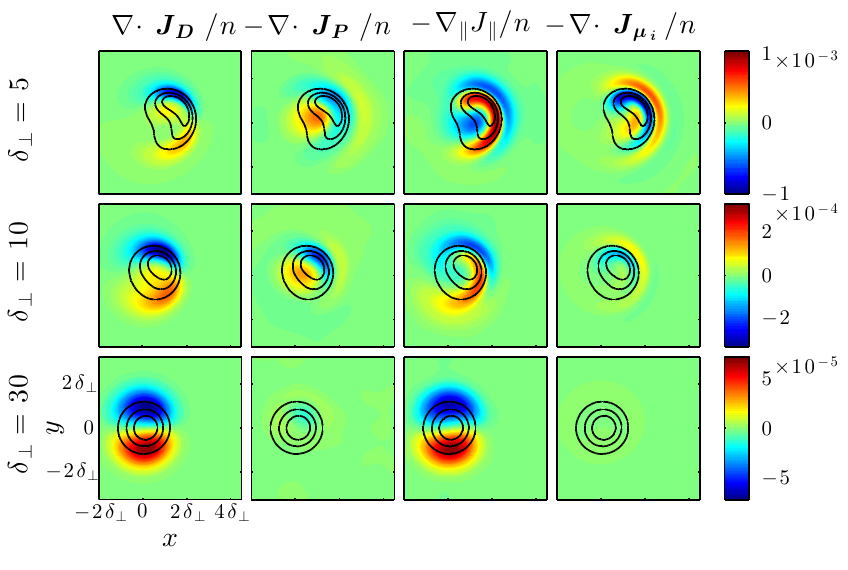}
\caption{Example magnitudes of the divergence of each current divided by $n$ at the midplane ($z = 0$) for $\delta n = 2n_0$, $L_b = L_\parallel/2$ filaments seeded with $\delta_\perp = 5$, $\delta_\perp = 10$ and $\delta_\perp = 30$.  The fields are shown at an example time $t = 250$ and density contour lines are also shown for reference.  }
\label{fig:current_balance}
\end{figure}

A greater understanding of the mechanism of this $\delta_\perp$ dependence can gained by determining which current paths are dominant in closing the non divergence free diamagnetic currents.  This can be inferred from Figure \ref{fig:current_balance}, %
which plots the magnitude of each of the terms in Equation \eqref{eq:vort} at the midplane ($z = 0$) for each $\delta_\perp$ filament at an example time early in their evolution, $t = 250$.  Density contour lines are also plotted for reference, and it is evident that the larger $\delta_\perp = 30$ filament's current balance is achieved almost entirely between $\boldsymbol{J_D}$ and $J_\parallel$ whilst $\boldsymbol{J_P}$ and $\boldsymbol{J_{\mu_i}}$ are both negligible.  In contrast, a more complicated current balance is observed in the two smaller filaments.  For the $\delta = 10$ filament, $J_\parallel$ predominantly closes $\boldsymbol{J_D}$ in the bottom lobe, whereas in the top lobe, $\boldsymbol{J_P}$ and $\boldsymbol{J_{\mu_i}}$ combine to ensure current continuity.  Finally, in the smallest $\delta_\perp = 5$ filament, $J_\parallel$ and $\boldsymbol{J_{\mu_i}}$ terms are the two dominant terms, but largely cancel each other out, with the remainder of their summed fields combining with the polarisation current term to close the non divergence free component of  $\boldsymbol{J_D}$.

\begin{figure}
\centering
\includegraphics[trim = 0mm 0mm 0mm 0mm, clip, width = 8.5cm]{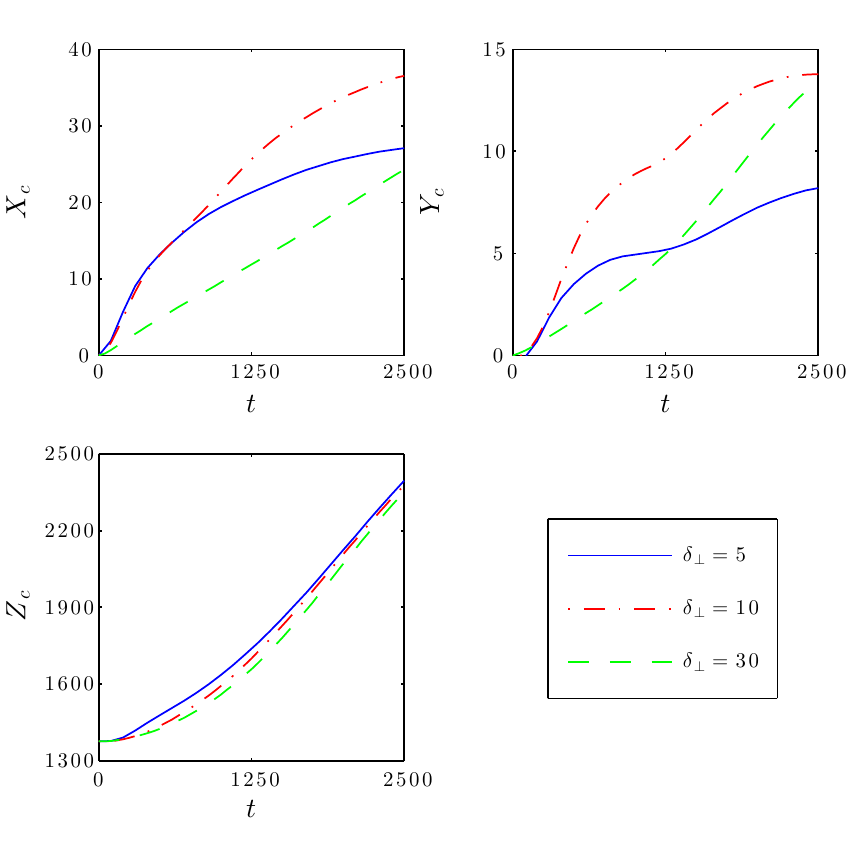}
\caption{Centre of mass coordinate evolution of $\delta n = 2n_0$, $L_b = L_\parallel/2$ filaments initialised with $\delta_\perp = 5$, $\delta_\perp = 10$ and $\delta_\perp = 30$.}
\label{fig:deltascan_cogs}	 
\end{figure}

A quantitative assessment of the net particle transport observed in each of these simulations can be achieved by calculation of the centre of mass coordinates of the density perturbation, $X_c$, $Y_c$, $Z_c$, defined as follows:
\begin{equation}
X_c =  \dfrac{\int n_b x\: d^3x}{\int n_b\: d^3x},\:  
Y_c =  \dfrac{\int n_b y\: d^3x}{\int n_b\: d^3x},\: 
Z_c =  \dfrac{\int n_b z\: d^3x}{\int n_b\: d^3x}.   
\label{eq:front_vs}  
\end{equation}
The evolution of these quantities for each $\delta_\perp$ is presented in Figure \ref{fig:deltascan_cogs}. %
The net radial motion of the two smallest filaments are approximately equal until $t = 800$, at which point the smaller $\delta_\perp = 5$ blob loses its coherence and decelerates more quickly, meaning that the $\delta_\perp=10$ filament exhibits a larger radial displacement at the end of the simulation, corresponding to a distance of approximately 6 cm.  The largest filament's initial net radial velocity is less than that of the others due to the fact that only the finger structure displays significant motions whilst the remainder of the initial blob is largely motionless.  Unlike the two smaller filaments, its net radial velocity remains largely constant throughout the simulation as the front of the finger retains its coherence because it is continually replenished due to the radial $\boldsymbol{E} \times \boldsymbol{B}$ velocity along the finger's column.  All three filaments exhibit a net poloidal displacement resulting from Boltzmann spinning effects, whilst their net parallel displacements are approximately equal throughout. 

\begin{figure}
\centering
\includegraphics[trim = 0mm 0mm 0mm 0mm, clip, width = 8.5cm]{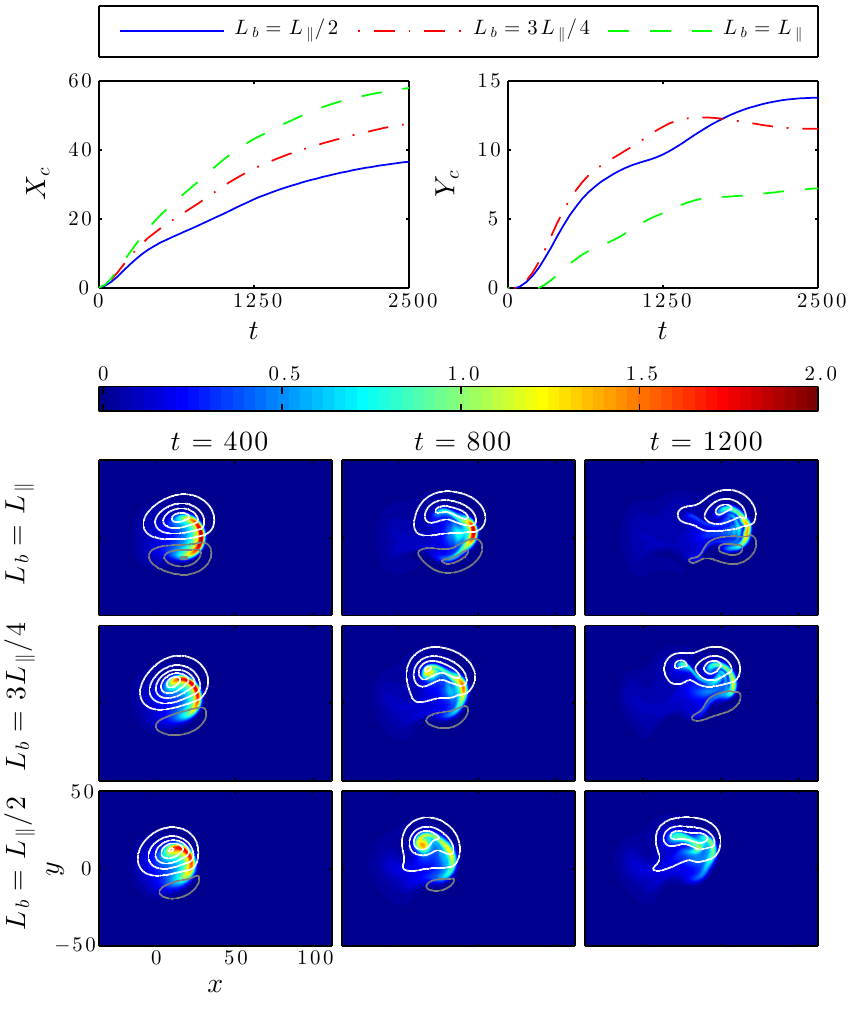}
\caption{Contrasting dynamics of filaments initialised with varying $L_b$.  All filaments initialised with $\delta n = 2n_0$ and $\delta_\perp = 10$.  Top: Evolution of perpendicular centre of mass coordinates.  Bottom: Midplane (z=0) Evolution of $n_b$ (colour map) and $\phi$ (contour lines) fields.  Grey and white contour lines respectively indicate values of $\phi$ less than and greater than the equilibrium value on the drift plane.} 
\label{fig:Lbscan}	 
\end{figure}

The influence of the filament's parallel extent has also been investigated by simulating filaments as in the reference case but with $L_b = 3L_\parallel/4$ and $L_b = L_\parallel$.  The results of this scan are shown in Figure \ref{fig:Lbscan}, %
which plots the evolution of the perpendicular centre of mass coordinates and also example time slices of the $n_b$ and $\phi$ fields at the midplane. A clear trend is evident in that the larger the parallel extent of the filament, the greater its net radial velocity throughout the simulation, resulting in a larger net radial displacement, $X_c$.  This trend is attributed to the enhanced Boltzmann response for shorter $L_b$, as this leads to the non divergence free diamagnetic current to be closed more through the $\boldsymbol{E}\times\boldsymbol{B}$ component of $\boldsymbol{J_P}$ in Equation \eqref{eq:vort} rather than directly through the $\frac{\partial\Omega}{\partial t}$ term.  This leads therefore to a reduction of the dipolar component of $\phi$ that produces radial advection, which can be qualitatively observed in the contour plots in Figure \ref{fig:Lbscan} by assessing the degree of asymmetry in the magnitude of the positive and negative poles of the $\phi$ structure.  

\begin{figure*}
\centering
\includegraphics[trim = 0mm 0mm 0mm 0mm, clip, width = 17.0cm]{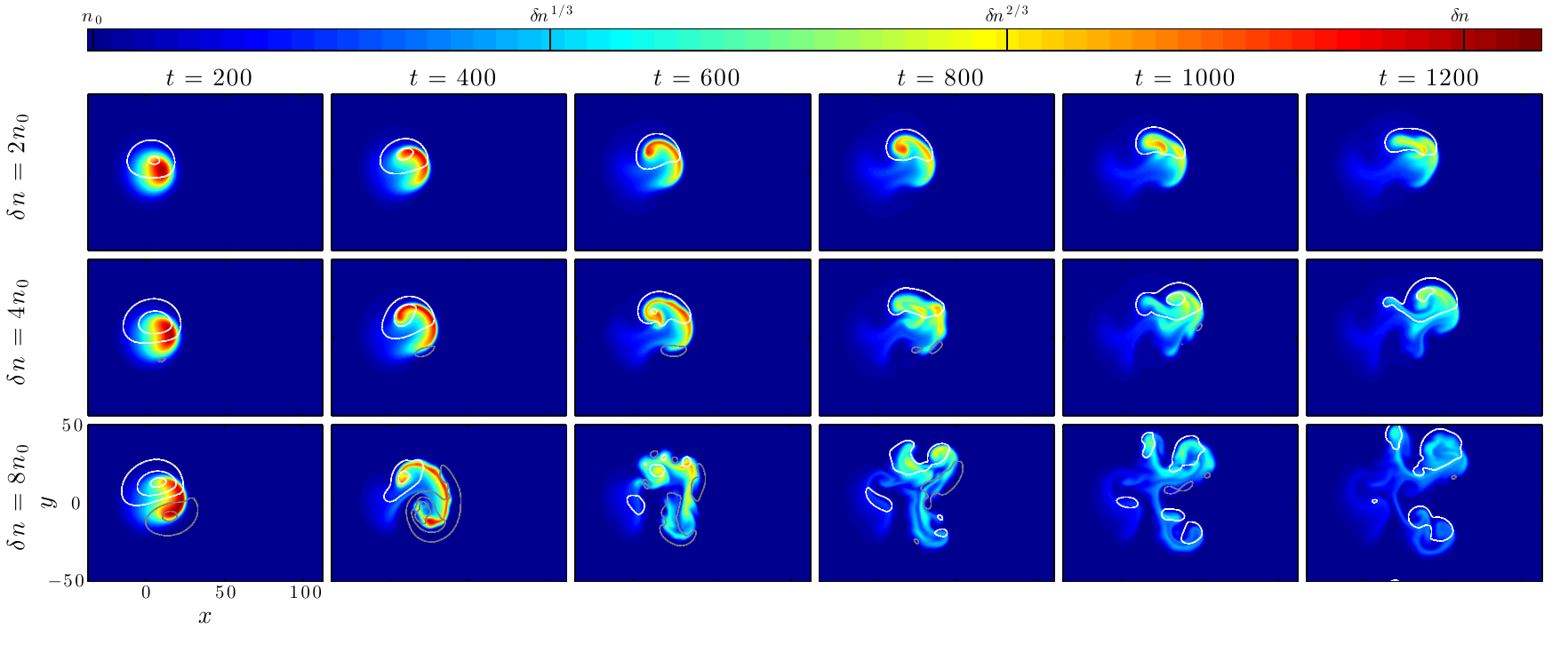}
\caption{Contrasting evolution of midplane ($z=0$) $n$ (colour map) and $\phi$ (contour lines) fields of filaments seeded with varying $\delta n$.  All filaments were otherwise initialised with $L_b = L_\parallel/2$ and $\delta_\perp = 10$.  White $\phi$ contours denote $\phi > \phi_{eq}(z = 0)$, whilst grey denotes $\phi < \phi_{eq}(z = 0)$.  The density fields are plotted using a logarithmic colour scale to resolve the filaments' structure late in the simulation.}
\label{fig:ampscan_conts}
\end{figure*}

\begin{figure}
\centering
\includegraphics[trim = 0mm 0mm 0mm 0mm, clip, width = 8.5cm]{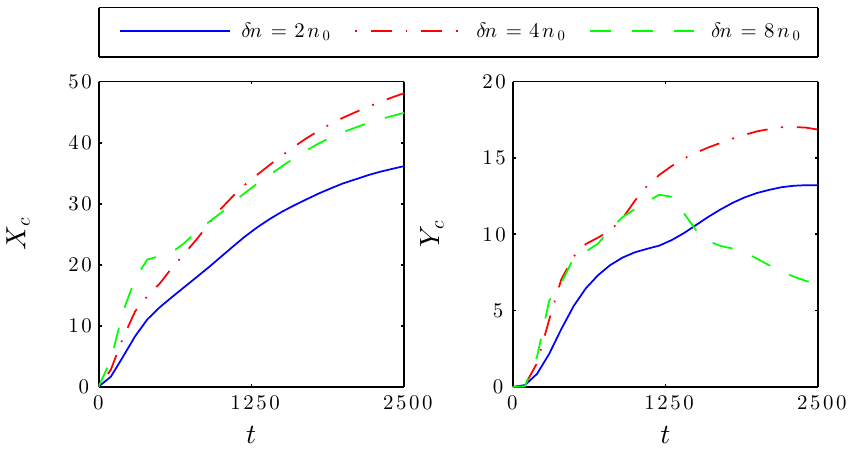}
\caption{Contrasting perpendicular centre of mass coordinate evolution of filaments with varying $\delta n$.  All filaments were otherwise initialised with $L_b = L_\parallel/2$ and $\delta_\perp = 10$.  }
\label{fig:ampscan_cogs}
\end{figure}

Next, the importance of the filaments' initial amplitude was studied, by simulating filaments with amplitudes twice and four times larger than in the reference case ($\delta n = 2n_0$).  The resolution was doubled for these higher amplitude cases to resolve the larger gradients that were produced.  The evolution of the midplane $n$ and $\phi$ fields of each of these simulated filaments in addition to the reference are shown in Figure \ref{fig:ampscan_conts} %
with the filled density contours plotted using a logarithmic colour scale to allow for the later density structures to be clearly visible.  In contrast to all of the simulations presented so far in this work, the highest amplitude filament exhibits unstable drift-wave dynamics, similar to those described in \cite{Angus:2012gk, Angus:2012fp}.  In those works, the onset of such dynamics was characterised by small density perturbations along the leading front of the filament in the perpendicular plane and along the parallel direction.  Such perturbations are observed at around $t = 400$ before the filament is subject to violent turbulent motions that act to tear it apart, destroying its coherence and rapidly dissipating its density.  Whilst the onset of such turbulence does inhibit the filament's net radial transport, it is not entirely halted as a number of smaller amplitude child filaments emerge from the turbulence which subsequently advect radially outwards themselves.  This effect can seen in Figure \ref{fig:ampscan_cogs}, %
which plots the contrasting evolution of $X_c$ and $Y_c$ for each of the filaments in this amplitude scan.  

\begin{figure}
\centering
\includegraphics[trim = 0mm 0mm 0mm 0mm, clip, width = 8.5cm]{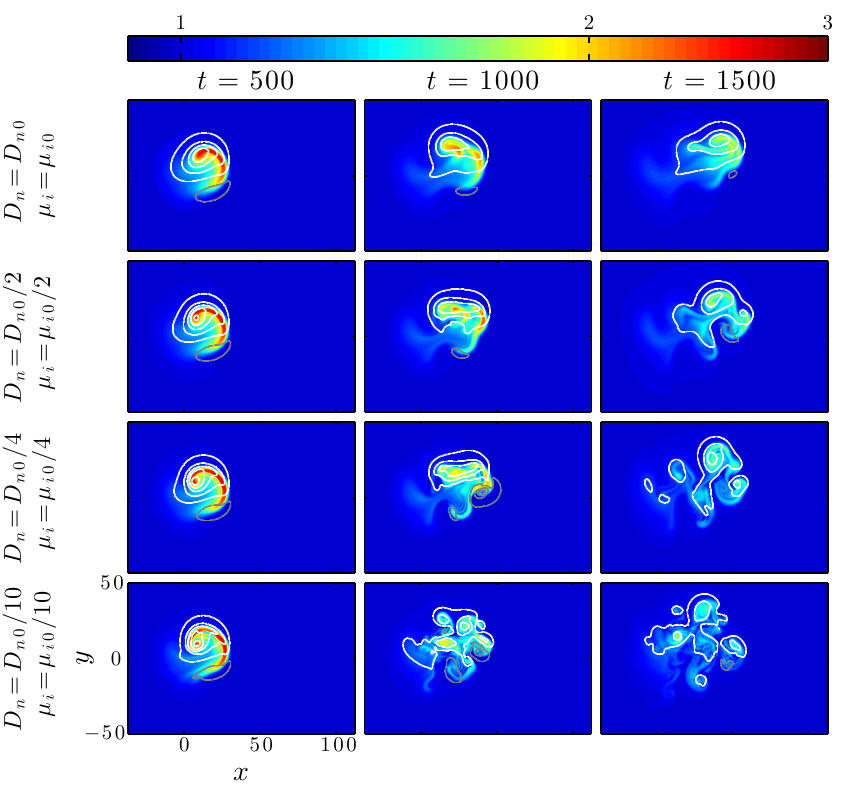}
\caption{Contrasting evolution of midplane ($z = 0$) $n$ (colour map) and $\phi$ (contour lines) of $\delta n = 2n_0$, $L_b = L_\parallel/2$, $\delta_\perp = 10$ filaments with varying dissipative parameters, $D_n$ and $\mu_i$.  Grey and white contour lines respectively indicate values of $\phi$ less than and greater than the equilibrium value on the drift plane.  The density fields are plotted using a logarithmic colour scale to resolve the filaments' structure late in the simulation. }
\label{fig:diss_scan}
\end{figure}

To determine the extent to which the dissipative parameters suppress instabilities in the lower initial amplitude cases, additional simulations were performed of $\delta n = 2$, $L_b=L_\parallel/2$, $\delta_\perp=10$ filaments with dissipative parameters reduced by factors of two, four and ten compared to the reference case values $D_{n0}$ and $\mu_{i0}$.  Example time slices from these results are plotted in Figure \ref{fig:diss_scan}, %
and it is clear that whilst the reference filament remains coherent and stable throughout its evolution, the filaments with smaller dissipative parameters break down into turbulence and produce a comparatively more diffuse filament by the final time frame.  When an equivalent scan in the dissipative parameters was also completed using the 2D sheath dissipation closure, the filaments remained largely coherent throughout, which suggests that the instability causing the filaments' breakdown in the bottom three rows of Figure \ref{fig:diss_scan} is a 3D specific effect.  

\begin{figure}
\centering
\includegraphics[trim = 0mm 0mm 0mm 0mm, clip, width = 8.5cm]{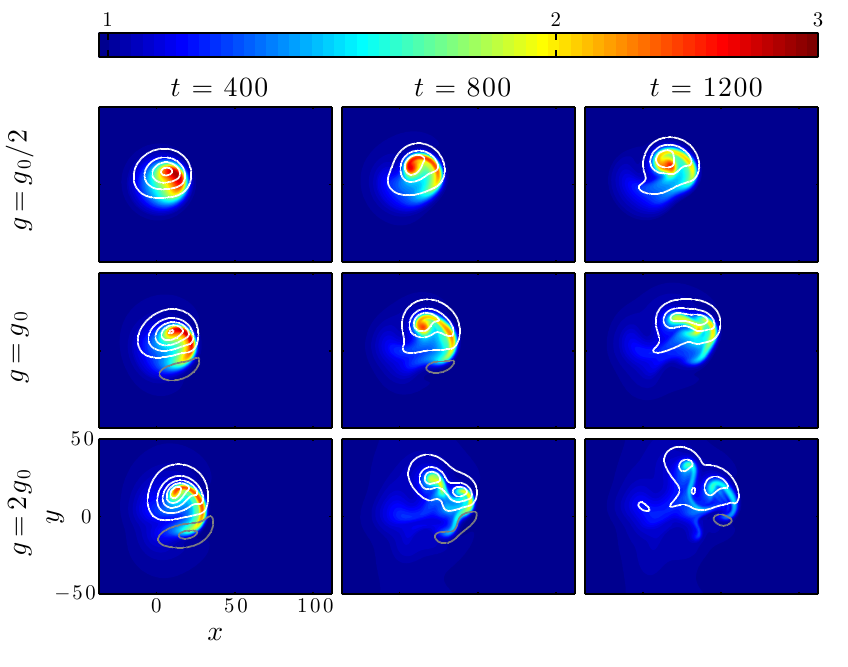}
\caption{Contrasting evolution of midplane ($z = 0$) $n$ (colour map) and $\phi$ (contour lines) of $\delta n = 2n_0$, $L_b = L_\parallel/2$, $\delta_\perp = 10$ filaments with varying curvature strength, $g$.  Grey and white contour lines respectively indicate values of $\phi$ less than and greater than the equilibrium value on the drift plane.  The density fields are plotted using a logarithmic colour scale to resolve the filaments' structure late in the simulation. }
\label{fig:gscan_conts}
\end{figure}

\begin{figure}
\centering
\includegraphics[trim = 0mm 0mm 0mm 0mm, clip, width = 8.5cm]{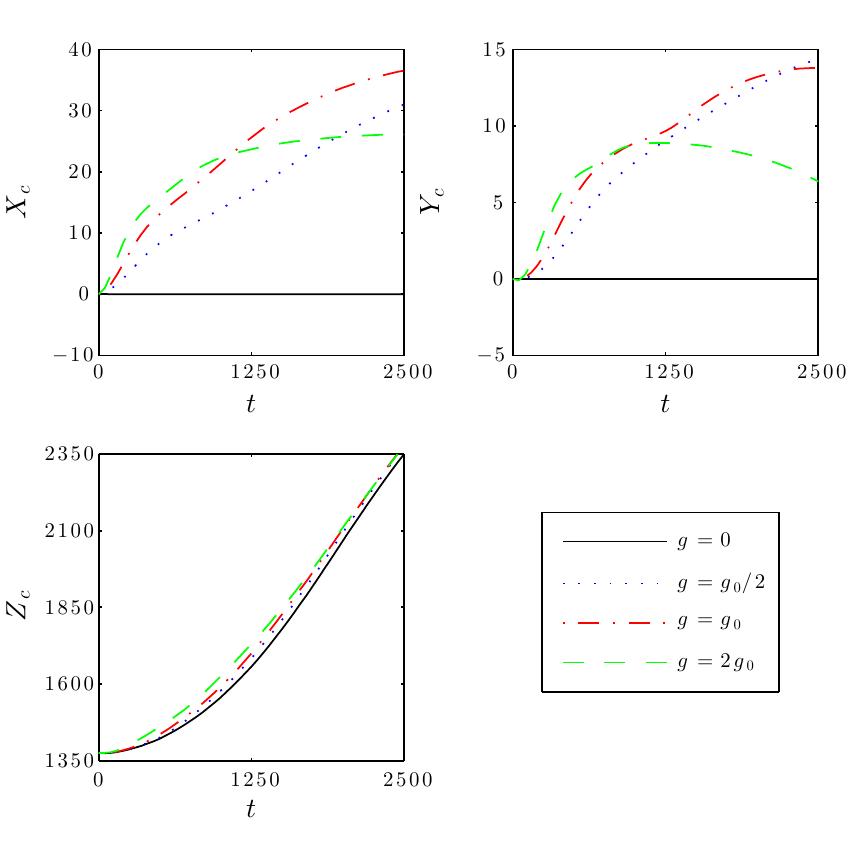}
\caption{Contrasting centre of mass coordinate evolution for $\delta n = 2n_0$, $L_b = L_\parallel/2$, $\delta_\perp = 10$ filaments with varying curvature strength, $g$.  }
\label{fig:gscan_cogs}
\end{figure}

To conclude the investigation into 3D filament dynamics, a scan in the strength of the magnetic curvature, $g$, was completed, with filaments seeded as in the reference case but with values of $g$ a factor of two smaller and larger than the MAST reference value, $g_0 = 0.0025$.  The midplane $n$ (colour map) and $\phi$ (contour lines) dynamics produced by each of these simulations are plotted in Figure \ref{fig:gscan_conts} %
with a logarithmic colour scale again employed for the density contours.  As the curvature is increased, the dipolar component of $\phi$ becomes stronger and this produces an increased initial radial velocity, which can be observed in Figure \ref{fig:gscan_cogs}, %
which displays the centre of mass evolutions with an with an additional $g = 0$ data series for comparison.  The $g = 2g_0$ filament thus forms a thin mushroom cap structure much faster than the reference $g = g_0$ filament, before producing a second mushroom structure from the peak density region of its first.  However, the filament's coherence is rapidly lost through these interchange motions in addition to collisional diffusion, and therefore its net radial displacement levels off at around $t = 1200$ in Figure \ref{fig:gscan_cogs}.  In contrast, the two filaments subject to smaller (but finite) curvature have undergone a greater net displacement.  In fact, the filament with $g = g_0/2$ retains its coherence throughout the simulation, as the Boltzmann spinning motions which enhances its togetherness appear to occur on a faster time scale than its interchange motions, and this is reflected in an approximately constant net radial velocity throughout.  Finally, in comparing the net motions of each of the filaments along the field line, it is observed that increasing curvature marginally increases $dY_c/dt$.  

In addition to the filament geometry studies described in detail in this section, the results were found to be insensitive to the values of $\nu_\parallel$ and $T_i$, with no significantly different dynamics observed for values an order of magnitude smaller or larger than the MAST reference case.  

\subsection{3D-1D Comparison}
\label{sec:1D3D}
\begin{figure}
\centering
\includegraphics[trim = 0mm 0mm 0mm 0mm, clip, width = 8.5cm]{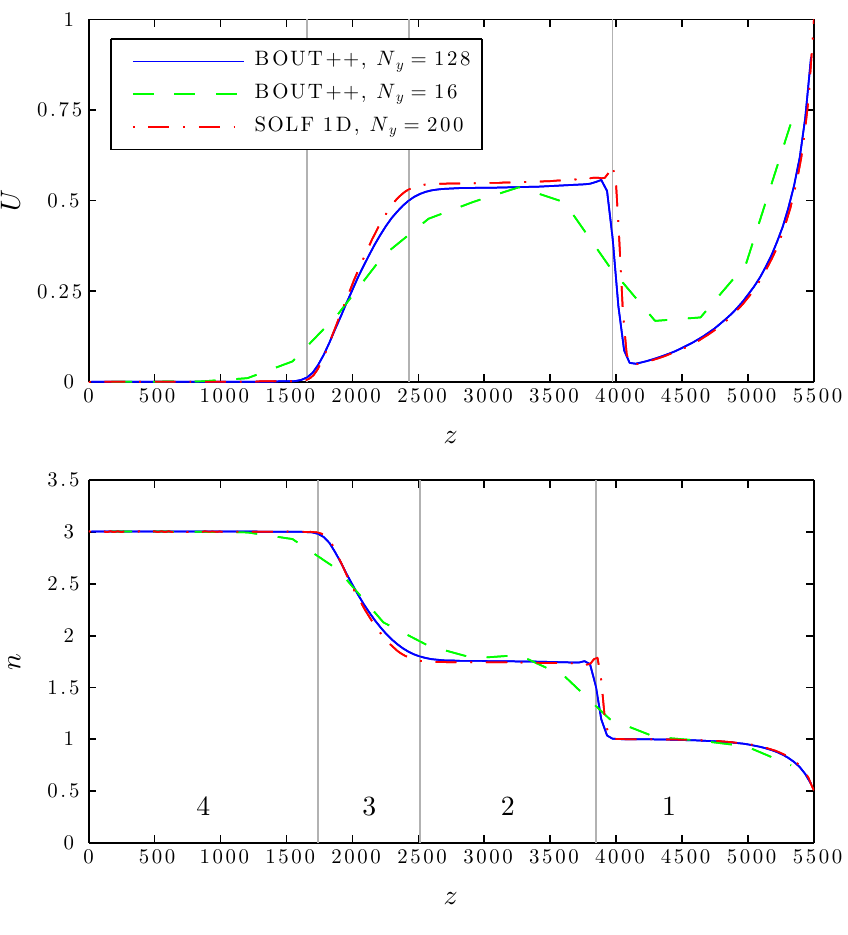}
\caption{Comparison of the parallel profiles of $U$ and $n$ produced by the 3D BOUT++ and 1D SOLF1D codes of a 1D filament seeded with $\delta n = 2n_0$, $\delta_\perp = \infty$, and $L_b = L_\parallel/2$ at an example time $t = 1000$. 
}
\label{fig:1D_profiles}
\end{figure}

The parallel dynamics of the BOUT++ code used for the 3D simulations outlined in the previous section have been compared against analytic theory and an existing SOL code.  For this verification exercise a 1D filament with $\delta n = 2n_0$, $\delta_\perp = \infty$ and $L_b = L_\parallel/2$ was simulated using a high resolution $N_y = 128$ mesh in addition to the default $N_y = 16$ mesh that is used throughout this paper.  The parallel motions of this 1D filament are representative of the drift plane averaged parallel dynamics observed in the 3D simulations presented in the previous section.  Example density and ion velocity profiles from these simulations at $t = 1000$ are shown in Figure \ref{fig:1D_profiles}, %
and it is noted that the electron velocity is approximately equal to its ion counterpart in this 1D case.  It is clear that particularly for the higher resolution simulations, four distinct regions exist in the parallel direction which are labelled in Figure \ref{fig:1D_profiles}.  At $t = 0$, only regions 1 and 4 exist, corresponding to the density of the background and in the region of the filament respectively.  As time progresses, a compressive shock moves into the lower density plasma, producing an area of intermediate $n$ and constant velocity, which is labelled region 2.  Simultaneously, a rarefaction wave, region 3, is created that moves in the opposition direction into the higher density plasma.  

The propagation of one dimensional shock fronts and rarefaction waves in a neutral gas is a well studied problem in fluid dynamics.  Since $U = V$ for this 1D case, the plasma can be treated as a single species fluid with velocity $U$, allowing the normalised speed of the interfaces shown in Figure \ref{fig:1D_profiles} %
\begin{figure}
\centering
\includegraphics[trim = 0mm 0mm 0mm 0mm, clip, width = 8.5cm]{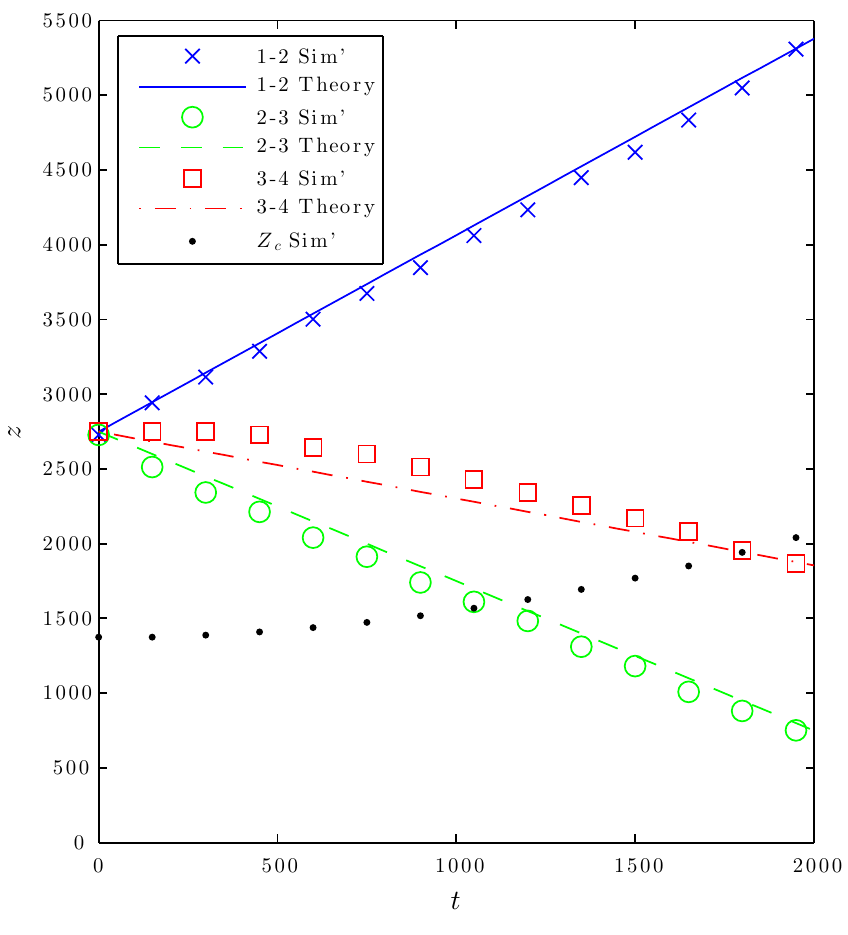}
\caption{Propagation of the interfaces of the regions labelled in Figure \ref{fig:1D_profiles} from the BOUT++ $N_y = 128$ simulation and comparison with analytical predictions of Equation \eqref{eq:front_vs}. }
\label{fig:1D_fronts}	 
\end{figure}
to be obtained from standard fluid shock theory \cite{Whitham:2011uz}:
\begin{equation}
 u_{12}=\sqrt{\dfrac{n_2}{n_1}},\;\;
 u_{23}=-1+\sqrt{\dfrac{n_2}{n_1}}-\sqrt{\dfrac{n_1}{n_2}},\;\; 
 u_{34}=-1,
\end{equation}
where $u_{ab}$ is the speed of the interface between Regions $a$ and $b$, and $n_a$ is the density in region $a$, and $n_2$ is found by numerically solving:
\begin{equation}
\ln{n_2} + \sqrt{\dfrac{n_2}{n_1}} - \sqrt{\dfrac{n_1}{n_2}} = \ln{ n_4}
\end{equation}
In Figure \ref{fig:1D_fronts}, the locations of these boundaries from the $N_y = 128$ simulation have been plotted versus time alongside the analytic predictions, and good agreement is found between simulation and theory for all three interfaces.  The position of the filament simulation's centre of mass in the parallel direction has also been plotted for reference.  

Figure \ref{fig:1D_profiles} also plots profiles obtained used the 1D SOL code SOLF1D \cite{Havlickova:2011dc} with $N_y = 200$ for comparison, and excellent agreement is found between SOLF1D and the higher resolution BOUT++ simulation.  In addition, the low resolution BOUT++ results are in reasonable agreement with the other two data series, despite evidently being more dissipative.  In particular, key features such as the density and velocity levels in each region are largely resolved, and justifies using $N_y = 16$ for the 3D simulations in this work.  

\subsection{3D-2D Comparison}
\label{sec:2D3D}
\begin{figure}
\centering
\includegraphics[trim = 0mm 0mm 0mm 0mm, clip, width = 8.5cm]{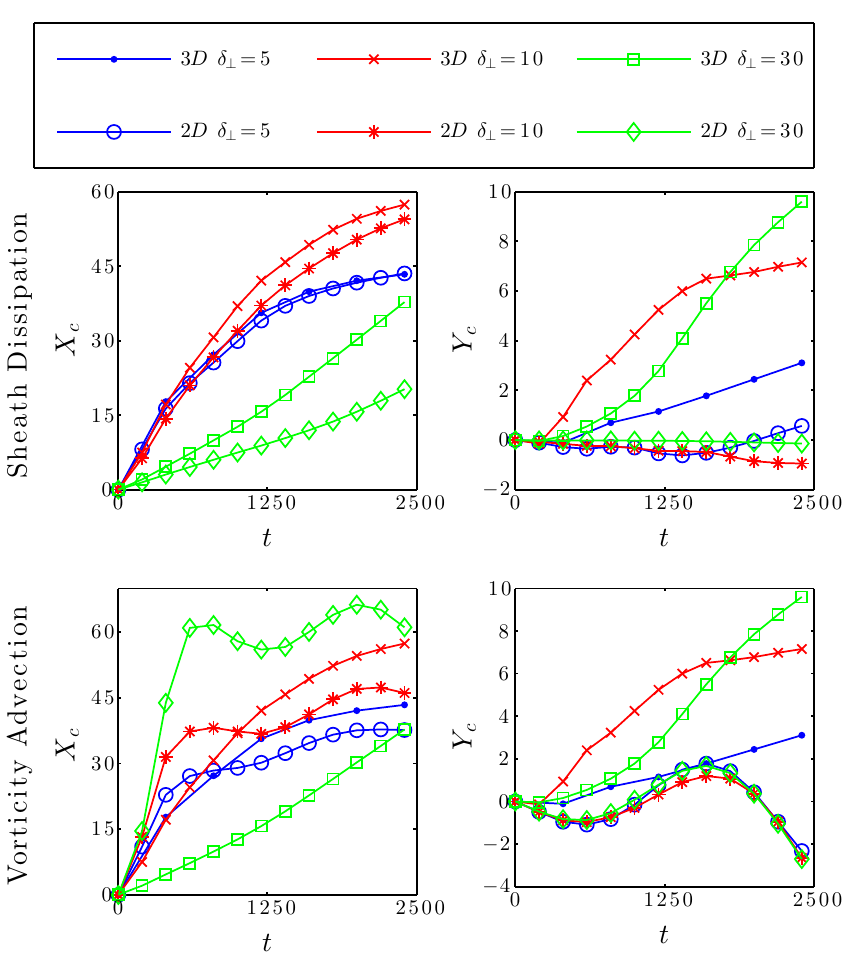}
\caption{Comparison of the perpendicular centre of mass coordinate evolution produced by the 2D sheath dissipation and vorticity advection closures against that from 3D simulations for filaments seeded with $L_b = L_\parallel$, $\delta n = 2n_0$ and a range of $\delta_\perp$.}
\label{fig:2D3D_Lb1_cogs}	 
\end{figure}

The remainder of the results presented in this work directly compare the 3D filament dynamics outlined in Section \ref{sec:3Dsims} with 2D simulations employing the sheath dissipation and vorticity advection closures.  Firstly, the ability of each of the 2D models to represent the sheath connected case can be assessed from Figure \ref{fig:2D3D_Lb1_cogs}, %
which plots the perpendicular centre of mass coordinate evolution obtained using all three models for filaments initialised with $\delta n = 2n_0$, $L_b = L_\parallel$ and values of $\delta_\perp$ equal to 5, 10 and 30.  The sheath dissipation model is most applicable for these cases, and the net radial motions obtained using this model, shown in the top left sub-plot, largely compare well with the 3D results.  In particular, excellent agreement is found for the smaller $\delta_\perp = 5$ and $\delta_\perp = 10$ filaments, although a less good comparison is found for the larger $\delta_\perp = 30$ filament, where parallel rather than polarisation currents are dominant in ensuring current continuity.  In contrast, the vorticity advection model overestimates the initial net radial velocity for all three filaments compared to the 3D simulations before they rapidly decelerate.  It can be noted from the two right sub-plots of the same figure that neither 2D model captures at all the net poloidal displacement observed in the 3D simulations, as neither has a drive by which $\phi$ can become monopolar and therefore no dipole rotation can occur.  

\begin{figure}
\centering
\includegraphics[trim = 0mm 0mm 0mm 0mm, clip, width = 8.5cm]{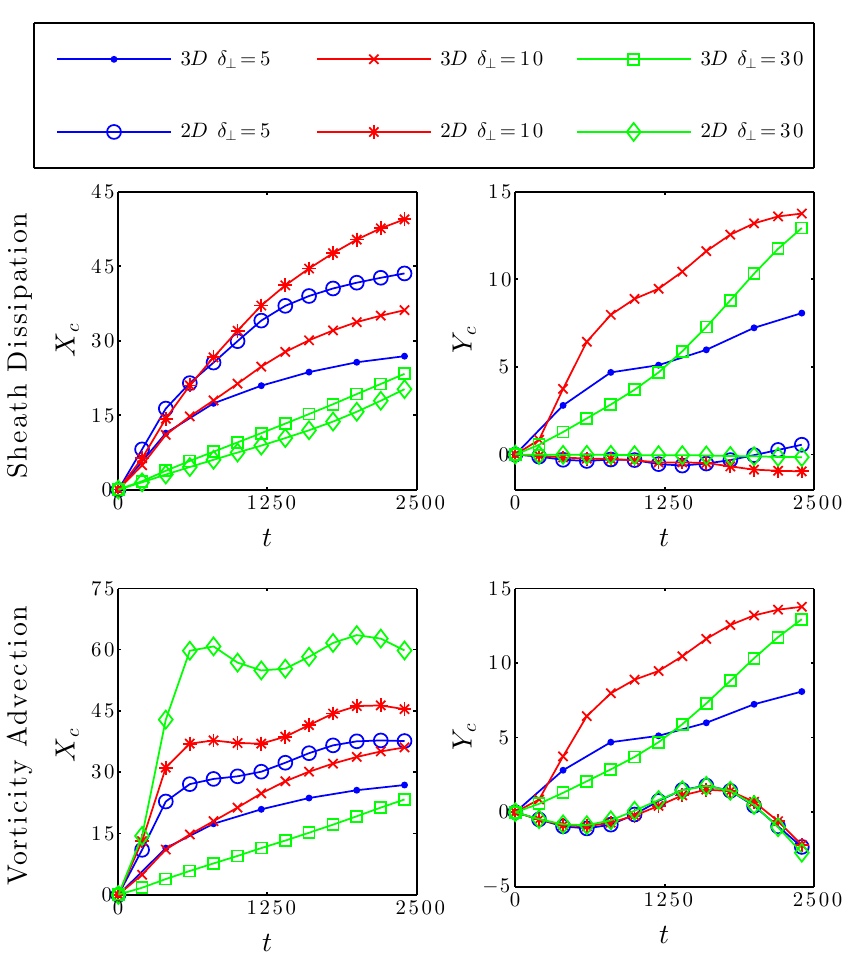}
\caption{Comparison of the perpendicular centre of mass coordinate evolution produced by the 2D sheath dissipation and vorticity advection closures against that from 3D simulations for filaments seeded with $L_b = L_\parallel/2$, $\delta n = 2n_0$ and a range of $\delta_\perp$.}
\label{fig:2D3D_Lb05_cogs}	 
\end{figure}

An equivalent investigation was also performed to determine the 2D models' performance in representing 3D filaments seeded with parallel gradients.  Figure \ref{fig:2D3D_Lb05_cogs} %
plots the same quantities as in Figure \ref{fig:2D3D_Lb1_cogs}, but for filaments initialised with $L_b = L_\parallel/2$ in the 3D and vorticity advection models.  Despite being constructed to be applicable for ballooned filaments, the vorticity advection model's net radial displacements again provide an unsatisfactory comparison with the 3D results.  On the other hand, whilst not being strictly justified in the presence of parallel gradients, the sheath dissipation closure's radial dynamics display better agreement with the 3D simulations, albeit not as satisfactory as in the $L_b = L_\parallel$ case.  In particular the contrasting dynamics observed for different $\delta_\perp$ produced by the 3D model are qualitatively reproduced.  In addition, by cross comparison between Figures \ref{fig:2D3D_Lb1_cogs} and \ref{fig:2D3D_Lb05_cogs}, it can be seen that the centre of mass evolutions produced by the vorticity advection are relatively insensitive to the value of $L_b$ with almost identical results yielded despite a factor 2 difference in $L_b$, although it is noted that the parallel draining of the blob was affected.  Therefore, for the remainder of this paper, the $L_b = L_\parallel/2$ case is taken as representative.

\begin{figure}
\centering
\includegraphics[trim = 0mm 0mm 0mm 0mm, clip, width = 8.5cm]{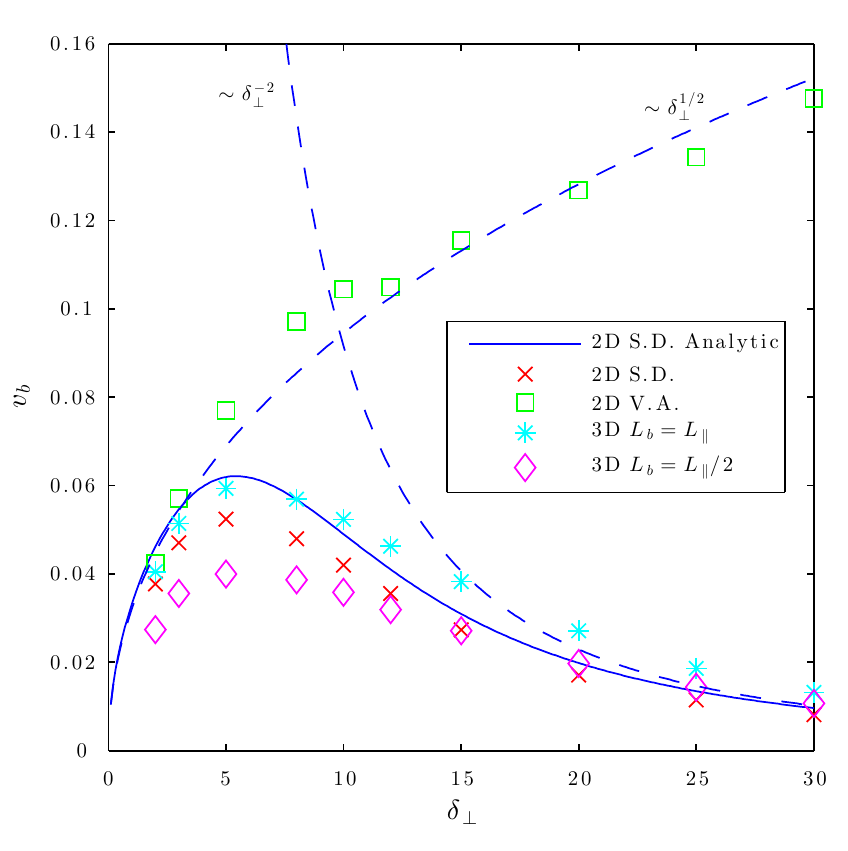}
\caption{Dependence of the characteristic radial velocity of the filaments during the early stages of motion on their initial $\delta_\perp$ for all 3 simulation models using filaments seeded with $\delta n = 2n_0$.  The sheath dissipation model's analytical scaling prediction of Equation \eqref{eq:vblob} is plotted for comparison.}
\label{fig:velocity_scaling}
\end{figure}

A distinctive feature of the filamentary motion produced by the vorticity advection model that can be observed from Figures \ref{fig:2D3D_Lb1_cogs} and \ref{fig:2D3D_Lb05_cogs} is that the filament's initial net radial velocity increases with $\delta_\perp$, which is in contrast to the results of the other two models.  This trend can be seen more clearly in Figure \ref{fig:velocity_scaling}, %
which shows the $\delta_\perp$ dependence of the characteristic radial velocity of the blob during the early phases of its motion, $v_b$, produced by each model for filaments seeded with $\delta n = 2n_0$.  Such a characteristic velocity was obtained from the simulations' output by selecting the peak value of $dX_c/dt$ during the first 20\% of the simulation.  The sheath dissipation scaling law of Equation \eqref{eq:vblob} is also plotted for comparison, and qualitative agreement is found with the simulations of the same model, with the velocity scaling like $\sim\delta_\perp^{1/2}$ for smaller $\delta_\perp$, and $\sim\delta_\perp^{-2}$ at larger $\delta_\perp$.  Crucially, a similar trend is observed for both the  $L_b = L_\parallel$ and the $L_b = L_\parallel/2$ cases of the 3D model.  The vorticity advection model however produces a characteristic velocity that scales like $\sim \delta_\perp^{1/2}$ for all $\delta_\perp$ and does not exhibit the roll over at higher $\delta_\perp$.  

\begin{figure}
\centering
\includegraphics[trim = 0mm 0mm 0mm 0mm, clip, width = 8.5cm]{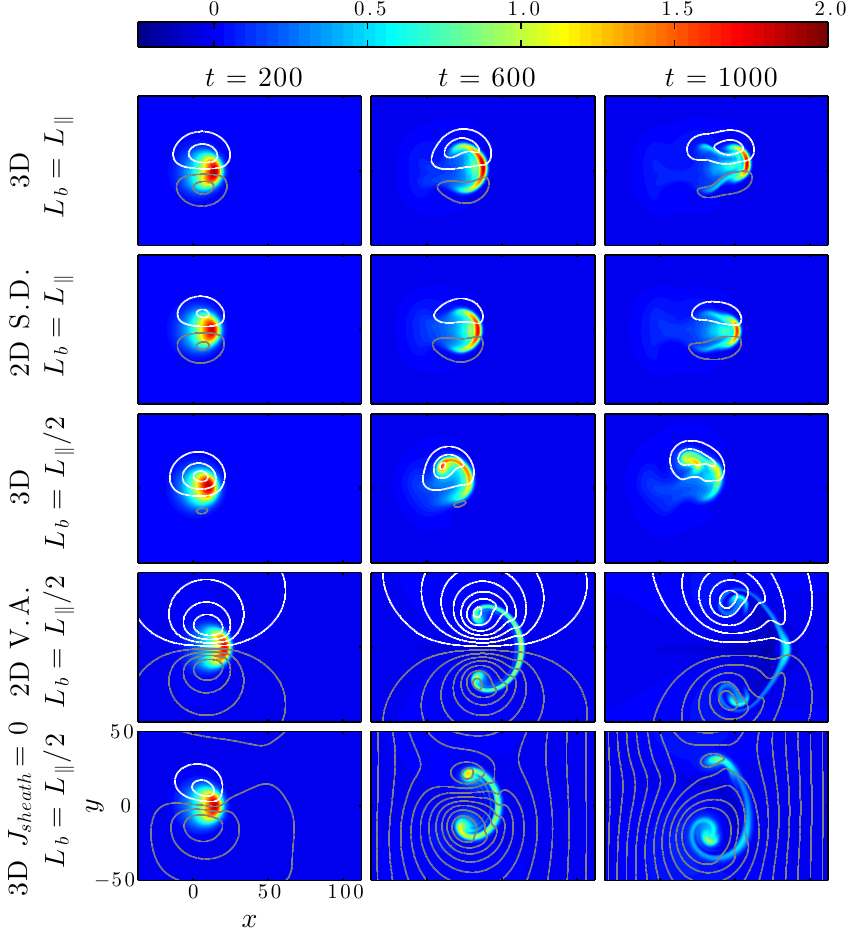}
\caption{Comparison of $n_b$ (colour map) and $\phi$ (contour lines) fields produced by the default 3D, sheath dissipation and vorticity advection models.  Additionally, the bottom row corresponds to a 3D simulation with $J_\parallel = 0$ enforced at the sheath.  The 3D fields are taken at the midplane ($z = 0$) and all filaments were seeded with $\delta n = 2 n_0$, $\delta_\perp = 10$ and $L_b$ as labelled.  Grey and white contour lines respectively indicate values of $\phi$ less than and greater than the equilibrium value on the drift plane.}
\label{fig:2D3D_conts}	 
\end{figure}

An insight into why the sheath dissipation model out performs the vorticity advection model in representing the 3D radial dynamics, even in the presence of parallel density gradients, can be obtained by contrasting the evolution of the density and potential fields produced by each model.  Such a comparison is provided in the top four rows of Figure \ref{fig:2D3D_conts}, %
which plots sample time slices of $n_b$ (colour map) and $\phi$ (contour lines) fields of all of the $\delta_\perp = 10$ simulations plotted in Figures \ref{fig:2D3D_Lb1_cogs} and \ref{fig:2D3D_Lb05_cogs}, with fields taken at the midplane for the 3D simulations.  From this, it is clear that the vorticity advection closure develops a much stronger dipole $\phi$ structure than the other two models, which corresponds to larger $E\times B$ flows and hence faster initial velocities.  In addition, the potential field extends further beyond the locality of the density perturbation, and this, combined with the faster velocities, produces larger $\boldsymbol{E}\times\boldsymbol{B}$ convective cells which act to rapidly stretch the leading front and mushroom the structure.  It is this mushrooming that corresponds to the rapid halts in net radial motion observed in Figures \ref{fig:2D3D_Lb1_cogs} and \ref{fig:2D3D_Lb05_cogs}.  

\begin{figure}
\centering
\includegraphics[trim = 0mm 0mm 0mm 0mm, clip, width = 8.5cm]{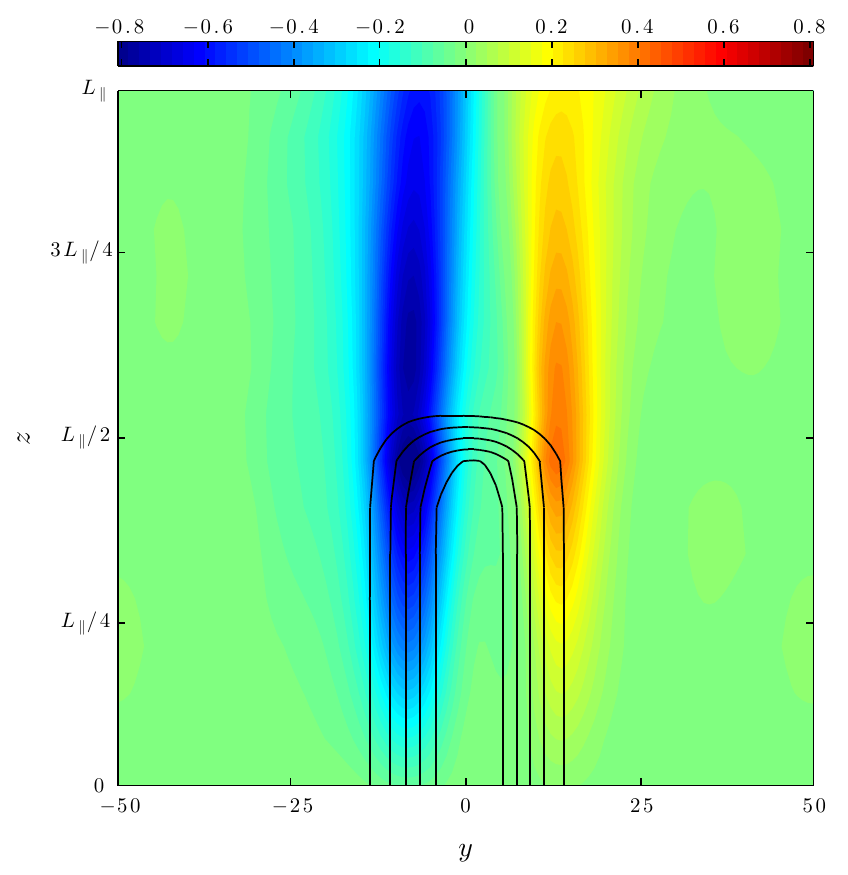}
\caption{Example $J_\parallel$ (colour map) structure in the $yz$ plane during the early phases of a $\delta n = 2 n_0$, $\delta_\perp = 10$, $L_b = L_\parallel$ simulation.  Density contour lines are also plotted for reference.  }
\label{fig:current_ex}	 
\end{figure}

In comparing the vorticity advection model directly to the sheath dissipation model, the qualitatively different potential fields arise due to the preferential damping of larger $\phi$ scale lengths by the sheath current term in Equation \eqref{eq:vort_SD}.  Therefore the apparent similarity between the sheath dissipation and 3D models indicate that parallel current paths completed through the sheath play a significant role in closing the non divergence free diamagnetic currents in the 3D simulation, especially for larger $\delta_\perp$.  The importance of parallel currents is immediately apparent from Figure \ref{fig:current_balance} and is further verified by Figure \ref{fig:current_ex}, %
which plots $J_\parallel$ (colour map) and density contours (contour lines) in an $yz$ plane in the region of the filament early in the evolution of the 3D $L_b = L_\parallel/2$ simulation.  It is clear that parallel currents are being closed through the sheath even when the filament's parallel front is far away from the sheath at the target.  Therefore the vorticity advection model's assumption that parallel currents are negligible in deriving Equation \eqref{eq:vort_VA} is clearly not applicable for this particular 3D model.  However, such a 3D model may not be universally valid and physically relevant situations may exist where increased resistivity in the region near the targets may prevent parallel currents from closing through the sheath.  Alternatively, enhanced magnetic shear near the X points may have a similar effect.  For example, in flux tube simulations using realistic magnetic geometry \cite{Walkden:2013dm}, it was observed that changing the boundary conditions at the target did not affect the filament's dynamics at the midplane.  

\begin{figure}
\centering
\includegraphics[trim = 0mm 0mm 0mm 0mm, clip, width = 8.5cm]{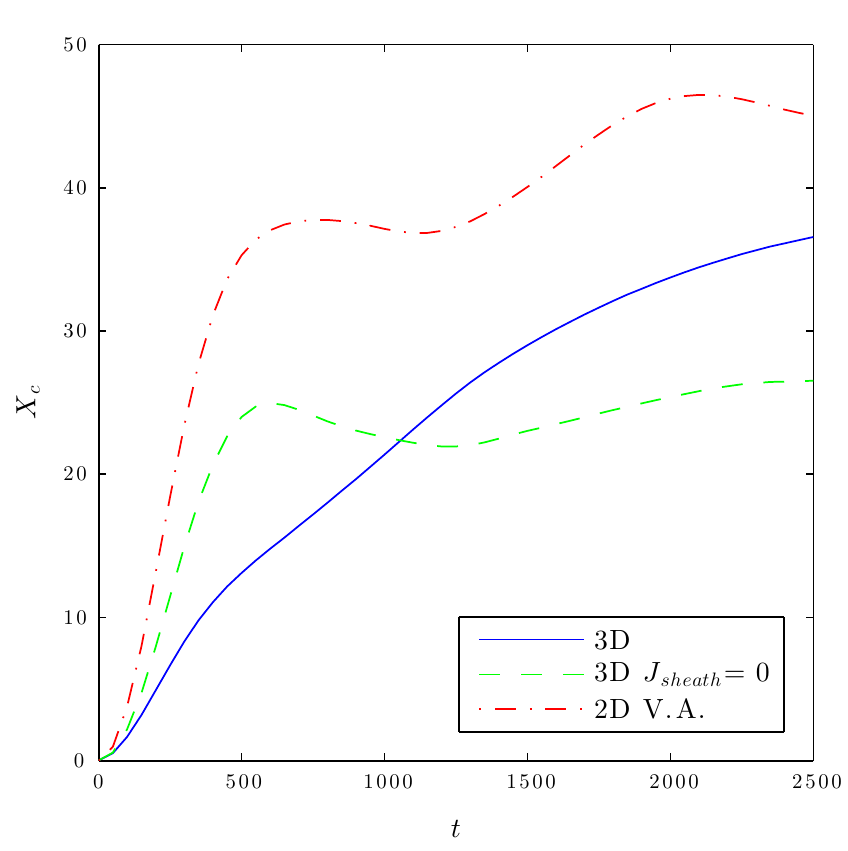}
\caption{Comparison of the radial centre of mass coordinate evolution produced by the default 3D model, the 3D model with $J_\parallel = 0$ enforced at the sheath and the vorticity advection model for a filament seeded with $\delta n = 2n_0$, $\delta_\perp = 10$, $L_b = L_\parallel/2$}.  
\label{fig:3Dnosheath_cogs}	 
\end{figure}

To further demonstrate the effect of sheath currents in the 3D simulation and to illustrate the effect of their inhibition, a further 3D $\delta n =2n_0$, $L_b = L_\parallel/2$, $\delta_\perp = 10$ simulation was performed, but with zero parallel current enforced at the sheath boundary.  Sample density and potential fields produced by this $J_{sheath} = 0$ simulation are plotted in the final row of Figure \ref{fig:2D3D_conts} and the results are qualitatively similar to those obtained using the vorticity advection model.  This similarity can also be seen Figure \ref{fig:3Dnosheath_cogs}, %
which compares the evolution of $X_c$ from this simulation with that produced by vorticity advection model and the default 3D model.  Clearly the absence of sheath currents in the 3D model leads to a much faster initial radial velocity, similar to that of the vorticity advection model, as stronger polarisation currents are driven in the absence of parallel currents.  

\begin{figure}
\centering
\includegraphics[trim = 0mm 0mm 0mm 0mm, clip, width = 8.5cm]{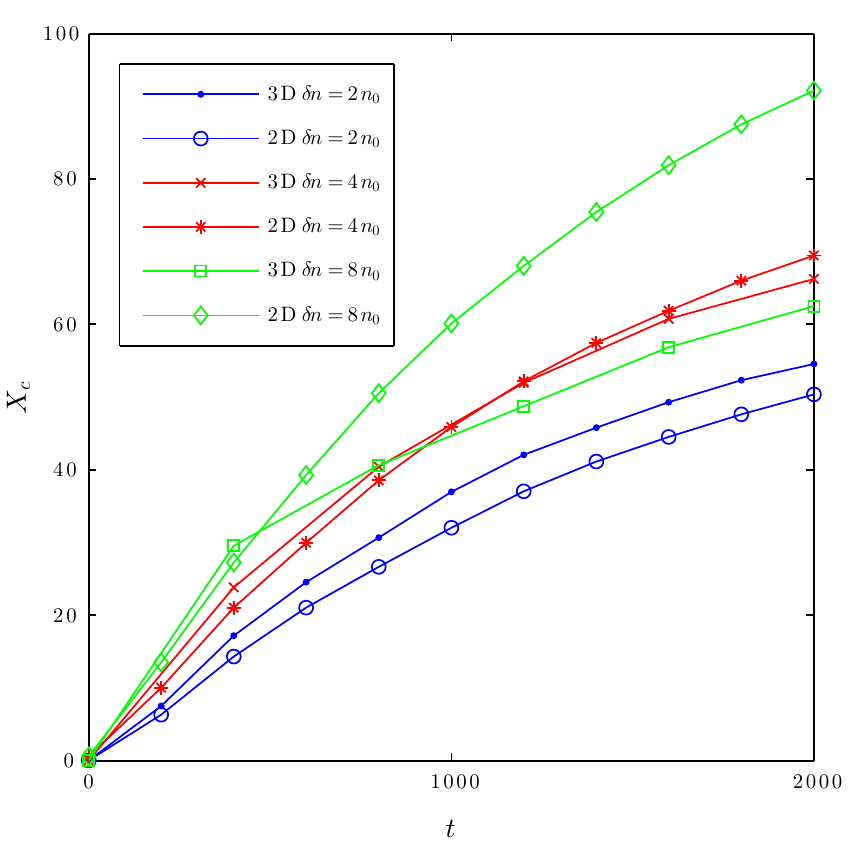}
\caption{Comparison of the radial centre of mass coordinate evolution produced by the 3D and sheath dissipation model for filaments seeded with $\delta_\perp = 10$, $L_b = L_\parallel/2$ and and varying $\delta n$.}
\label{fig:2D3D_amp_cogs}	 
\end{figure}

Finally, it is important to emphasise that the sheath dissipation model only accurately captures the 3D radial particle transport in the absence of drift-wave dynamics in the 3D simulations.  This is demonstrated in Figure \ref{fig:2D3D_amp_cogs}, %
which compares the 3D and sheath dissipation evolution of $X_c$ for filaments initialised with $L_b = L_\parallel$, $\delta_\perp = 10$ and varying $\delta n$.  The sheath dissipation model accurately matches the 3D results for $\delta n = 2n_0$ and $\delta n = 4n_0$, and in the early stages of $\delta n = 8n_0$ until $t \approx 500$. At this point however, drift-wave turbulence develops in the 3D simulation significantly reducing its radial transport.  Such drift-wave turbulence cannot be represented in the 2D model, and so it subsequently significantly overestimates $X_c$ after its onset in the 3D simulation.    

\section{Conclusions}
\label{sec:conclusions}
In this work the dynamics of isolated filament structures have been numerically simulated using a 3D model utilising sheath boundary conditions in the parallel direction, with filaments seeded onto a self-consistent source driven background.  The initial perpendicular size of the filament, $\delta_\perp$, was found to have a strong influence on its subsequent motion, as it determined whether current paths other than the parallel currents were important in closing the non divergence free diamagnetic current.  Smaller filaments, where polarisation and viscous currents did play a role, were observed to undergo mushrooming motions as a whole in the perpendicular plane.  Filaments with larger $\delta_\perp$ on the other hand, moved initially in the drift plane not as a single entity, but instead ejected a finger like structure whose leading front underwent similar interchange dynamics to those of the smaller filaments.  In the presence of parallel density gradients, the filaments' perpendicular $\phi$ fields were found to develop a monopolar component due to a Boltzmann response along the field line which induced these filaments to spin in the drift plane.  This Boltzmann spinning was observed to rotate the dipolar component of $\phi$ and therefore induce the filaments to display a net poloidal displacement.  Moreover, as the parallel extent of the initial filament was reduced, its net radial displacement and velocity also decreased, due to an enhancement of this Boltzmann response.  The majority of filaments were found to be stable to drift-wave instabilities due to dissipative effects, although drift-wave turbulence was observed when the filament's initial amplitude was sufficiently large.  As the values of the dissipative parameters used for these simulations are physically justified, this indicates that some filaments may exist in the MAST SOL that are drift-wave stable.  
      
The 3D simulations were compared to two 2D closures used prominently in the literature, namely, the sheath dissipation closure, which neglects parallel gradients, and the vorticity advection closure, which neglects the influence of parallel currents.  The vorticity advection closure was found to not replicate the 3D perpendicular dynamics well and overestimated the initial radial velocity of all the filaments studied.  In contrast, more satisfactory comparisons with the sheath dissipation closure's radial motions were obtained, even for 3D filaments with significant parallel gradients, where the closure is no longer strictly valid.  Specifically it captured the contrasting dynamics of filaments with different perpendicular sizes that were observed in the 3D simulations, which the vorticity advection closure failed to replicate.  However neither 2D closure was found to replicate the alignment of $n$ and $\phi$ and the associated net poloidal displacement of the filament upwards that was observed in the 3D simulations.  Moreover, the sheath dissipation model only accurately captured the 3D radial particle transport in the absence of drift-wave dynamics in the 3D simulations.  

It is concluded that the sheath dissipation closure was more successful in replicating the 3D dynamics, because non-negligible parallel currents that closed through the sheath were observed to occur in 3D simulations, even for filaments localised in the parallel direction.  However, it is possible that the vorticity advection closure may still be relevant for situations where currents are inhibited from closing through the sheath due to increased resistivity in the locality of the targets or due to enhanced magnetic shear around the X-point region.



\begin{acknowledgments}
L.E would like to acknowledge useful discussions with N. Walkden and J. Madsen.  This project has received funding from the European Union's Horizon 2020 research and innovation programme under grant agreement number 633053 and from the RCUK Energy Programme [grant number EP/I501045]. To obtain further information on the data and models underlying this paper please contact PublicationsManager@ccfe.ac.uk*. The views and opinions expressed herein do not necessarily reflect those of the European Commission.  In addition, this work was carried out also using the Plasma HEC Consortium EPSRC grant number EP/L000237/1 and the HELIOS supercomputer system at Computational Simulation Centre of International Fusion Energy Research Centre (IFERC-CSC), Aomori, Japan, under the Broader Approach collaboration between Euratom and Japan, implemented by Fusion for Energy and JAEA.
\end{acknowledgments}

\bibliography{Library.bib}

\end{document}